\documentclass[11pt]{article}

\usepackage[utf8]{inputenc}
\usepackage[T1]{fontenc}
\usepackage{lmodern}

\usepackage{amsmath}
\usepackage{amssymb}
\usepackage{amsthm}

\usepackage{authblk}

\newtheorem{proposition}{Proposition}
\usepackage{graphicx}

\usepackage[colorlinks, citecolor=blue,urlcolor=blue]{hyperref}

\usepackage{fullpage}

\usepackage[linesnumbered,ruled,vlined]{algorithm2e}
\SetKwInput{KwInput}{Input}           % Set the Input
\SetKwInput{KwOutput}{Output}         % set the Output

\usepackage{booktabs}
\usepackage{changepage}
\usepackage[labelfont=bf]{caption}
\usepackage{subcaption}
\usepackage{multirow}
\usepackage{tikz}
\usepackage{placeins}

\usepackage{natbib}

\newcommand{\Gcal}{\mathcal{G}}
\newcommand{\Lcal}{\mathcal{L}}
\newcommand{\Mcal}{\mathcal{M}}

\def\R{\mathbb{R}} % Reals
\DeclareMathOperator{\Tr}{Tr}

\DeclareMathOperator*{\maximize}{maximize}
\DeclareMathOperator*{\minimize}{minimize}

\title{Smooth tensor decomposition with application to ambulatory blood pressure monitoring data}

\author[1]{Leyuan Qian}
\author[2]{R. Nisha Aurora}
\author[3]{Naresh M. Punjabi}
\author[1]{Irina Gaynanova\thanks{Corresponding author: \href{mailto:irinagn@umich.edu}{irinagn@umich.edu}}}

\affil[1]{Department of Biostatistics, University of Michigan, Ann Arbor, Michigan, U.S.A.}
\affil[2]{Grossman School of Medicine, New York University, New York, New York, U.S.A.}
\affil[3]{Miller School of Medicine, University of Miami, Miami, Florida, U.S.A.}

\date{}

\begin{document}

\maketitle

\begin{abstract}
Ambulatory blood pressure monitoring (ABPM) enables continuous measurement of blood pressure and heart rate over 24 hours and is increasingly used in clinical studies. However, ABPM data are often reduced to summary statistics, such as means or medians, which obscure temporal features like nocturnal dipping and individual chronotypes. Functional data analysis methods better capture these temporal dynamics but typically treat each ABPM measurement separately, limiting their ability to leverage correlations among matched measurements. In this work, we observe that aligning ABPM data along measurement type, time, and patient ID lends itself to a tensor representation---a multidimensional array. Although tensor learning has shown great potential in other fields, it has not been applied to ABPM data. Existing tensor learning approaches often lack temporal smoothing constraints, assume no missing data, and can be computationally demanding. To address these limitations, we propose a novel smooth tensor decomposition method that incorporates a temporal smoothing penalty and accommodates missing data. We also develop an automatic procedure for selecting the optimal smoothing parameter and tensor ranks. Simulation studies demonstrate that our method reliably reconstructs smooth temporal trends from noisy, incomplete data. Application to ABPM data from patients with concurrent obstructive sleep apnea and type 2 diabetes uncovers clinically relevant associations between patient characteristics and ABPM measurements, which are missed by summary-based approaches.
\end{abstract}


\maketitle

\section{Introduction}
\label{sec:intro}

Wearable devices are increasingly used to monitor human physiological parameters in free-living conditions, enabling real-time health assessment and supporting disease diagnosis and management. Among these, ambulatory blood pressure monitors (ABPMs) provide 24-hour measurements of blood pressure and heart rate while individuals maintain their usual daily routines, including during sleep \citep{pickering_ambulatory_2006}. Unlike conventional in-office blood pressure measurements, which capture only isolated time points and miss nocturnal patterns, ABPMs offer significant advantages by capturing diurnal variability and accounting for environmental and behavioral influences \citep{kario_expert_2019}. These features make ABPM data valuable for studying how individual characteristics influence blood pressure and heart rate, including their temporal dynamics. In our motivating application, we focus on patients with coexisting type 2 diabetes and obstructive sleep apnea (OSA), a common sleep disorder characterized by recurrent blockage of the upper airway. OSA is a known independent risk factor for hypertension and cardiovascular disease \citep{punjabi_epidemiology_2008}. We aim to characterize ABPM patterns in this population and examine their associations with demographic and clinical factors, including OSA severity. However, current analytical approaches for ABPM data often lack the sufficient sensitivity needed to detect clinically relevant patterns. 

Traditional analyses of ABPM data rely on summary statistics, such as means of the measurements, calculated over the entire 24-hour period or within specific time segments like designated daytime and nighttime \citep{parati_european_2014}. %, %torres_use_2015}. 
Such approaches may not fully leverage the continuous, dynamic nature of ABPM profiles and the potential correlation between ABPM measurements, such as systolic and diastolic blood pressure. Applying a uniform division of daytime and nighttime across individuals may also introduce confounding, since different individuals have different sleep times. In our motivating application, the use of summary statistics leads to insufficient statistical power to detect the associations between OSA severity and blood pressure and heart rate (Section~\ref{sec:data}). 

Functional data analysis (FDA) methods offer alternative approaches for analyzing ABPM data by explicitly modeling temporal patterns. Common techniques used with ABPM data include cosinor models \citep{ayala_circadian_1997, madden_morning_2018}, linear mixed effects model with polynomials \citep{edwards_analysis_2014}, spline smoothing \citep{lambert_analysis_2001, madden_exploring_2017, murden_associations_2025}, and functional principal component analysis (FPCA) \citep{madden_morning_2018}. 
Each of these methods has its own advantages and limitations. Most notably, they are primarily designed for single functional trajectories and do not naturally accommodate joint modeling of multiple time-varying measures. 
This limits their ability to characterize the joint behavior and temporal interplay among systolic and diastolic blood pressure and heart rate.

Tensors are generalizations of matrices for higher-order data, which naturally arise with imaging and multichannel EEG signals \citep{lu_survey_2011, bi_tensors_2021}. It has been shown that statistical methods that take advantage of the intrinsic tensor structure have higher accuracy than methods that collapse the tensor objects in the matrix form \citep{cichocki_tensor_2015}. We observe that the ABPM data can be represented as a three-way tensor (hour $\times$ measurement type $\times$ patient) (Figure~\ref{fig:GLRAM}). The first mode (hour) captures temporal information; the second mode (measurement type) captures cardiovascular measures, such as systolic blood pressure (SBP), diastolic blood pressure (DBP), and heart rate (HR); and the third mode includes patient IDs. This intrinsic data structure motivates our idea of adopting tensor learning for ABPM data. However, while
multiple unsupervised tensor compression methods have been
developed \citep{kolda_tensor_2009}, they possess certain limitations that restrict their application to ABPM data. In particular, the smoothness of ABPM trajectories over time and the presence of missing data are not adequately addressed by existing methods.  

CANDECOMP/PARAFAC (CP) decomposition expresses a tensor as a sum of rank-one components and is identifiable under mild conditions \citep{kolda_tensor_2009, bi_tensors_2021}. Tucker decomposition \citep{tucker_mathematical_1966} is more flexible than CP and gives rise to multilinear singular value decomposition (MLSVD) \citep{de_lathauwer_multilinear_2000}, providing means to extend principal component analysis to tensors. However, both CP and Tucker decompositions assume fully observed data and lack smoothness constraints. Nuclear norm minimization methods \citep{recht_guaranteed_2010}, originally developed for low-rank matrix approximation, have been extended to tensors and used for regularized estimation with missing data \citep{yuan_tensor_2016, friedland_nuclear_2017}. However, they involve complex optimization procedures and similarly do not address temporal smoothness. A few existing works consider smoothing with tensor data, but primarily focus on CP decomposition. \citet{roald_ao-admm_2022} consider a general optimization framework for CP decomposition with constraints, out of which a graph Laplacian regularizaiton could be utilized to incorporate smoothness based on second-order difference matrix, an approach briefly mentioned in Chapter~6.2 of \citet{bro_1998}.  More recently, \citet{larsen2024tensordecompositionmeetsrkhs} allow for continuous tensor models via connections with reproducing kernel hilbert spaces, also focusing on CP decomposition. However, CP decomposition is more restrictive than Tucker and can exhibit degeneracy and a lack of optimal solutions for small, low-rank tensors, such as ABPM \citep{krijnen_non-existence_2008, zhou_tensor_2021}. Tucker decomposition is more flexible and avoids these issues. To our knowledge, the only prior work incorporating smoothness in Tucker decomposition is \citet{timmermanThreewayComponentAnalysis2002}, but it relies on pre-specification of smooth bases and does not enforce orthogonality across components, complicating interpretability and identifiability.

Motivated by the characteristics of ABPM data, we introduce a new smooth tensor decomposition method which builds on the Tucker decomposition but incorporates a temporal smoothing penalty. In contrast to existing work on smoothing with tensors, we penalize the reconstructed fit rather than the component matrices. This distinction reflects a more realistic modeling assumption—namely, that the data (not the components) are smooth—and is crucial for maintaining model stability. In our empirical studies, enforcing both smoothness and orthogonality on components led to poor fit and instability, a conflict that has not been identified in prior work (\ref{sec:smooth_on_L}). We develop an efficient iterative algorithm to solve the resulting optimization problem, extend it to accommodate missing data, and design a tailored cross-validation procedure to automatically select the ranks and smoothing parameter by masking observed entries to mimic missingness. To address the inherent non-uniqueness of Tucker decomposition, we apply SVD-based rotations of tensor unfoldings to ensure identifiability within MLSVD framework \citep{de_lathauwer_multilinear_2000}. The hour mode components capture temporal features, while those from the measurement mode summarize patterns across measurement types. The resulting core tensor scores can be used for downstream inference. 
In simulations, our method outperformed FPCA in recovering both overall signal and temporal components (Section~\ref{sec:simu}). In application, the method isolated distinct signals corresponding to overall blood pressure and heart rate levels, nocturnal dipping, and individual sleep time differences (Section~\ref{sec:data}). The achieved separations allowed for isolating the level-related signal for regression analysis without manually aligning patients by sleep time. This enabled us to detect a clinically meaningful association between OSA severity and overall blood pressure and heart rate—an effect that was not identified using traditional summary measures—demonstrating the increased statistical power from our approach. 

In summary, our main contributions are: (i) a novel method for low-rank tensor decomposition with temporal smoothing, (ii) an efficient optimization algorithm for fitting the decomposition that also allows for missing data, (iii) an automatic hyperparameter tuning procedure tailored for tensor data, (iv) application to ABPM data from patients with OSA that provides interpretable representations of main directions of variability
and better power due to information borrowing. While motivated by ABPM data, the method is broadly applicable to other settings involving temporally structured matched multivariate data.

\textbf{Notation:} We let $M\in \R^{n \times p}$ be a $n\times p$ matrix and $\|M\|_F$ its Frobenius norm.  Given a set of indices $\Omega$, we let $\|M\|_{\Omega} =\sqrt{ \sum_{(i,j)\in \Omega}m_{ij}^2}$ be the Frobenius norm restricted to that set. 
For two matrices $A$ and $B$ with the same dimensions, we let $A \circ B$ be the elementwise product. We let $\Mcal = (m_{i_1 i_2 \ldots i_k}) \in \mathbb{R}^{I_1 \times I_2 \times ... \times I_K}$ be a $K$th-order tensor, an array with $K$ dimensions (ways or modes), each of size $I_k$, $k=1, \dots, K$, with elements $m_{i_1 i_2 \ldots i_k}$.  We let $|\mathcal{M}|$ denote the total number of elements in $\Mcal$, and $M_{(k)}  \in \mathbb{R}^{I_k \times (I_1I_2 \ldots I_{k-1}I_{k+1}\ldots I_K)}$ be the mode-$k$ unfolding, a rearrangement of the tensor into a matrix along $k$th mode. Let $\|\Mcal\|_F = \sqrt{\sum_{i_1=1}^{I_1} \sum_{i_2=1}^{I_2} \ldots \sum_{i_K=1}^{I_K} (m_{i_1 i_2 \ldots i_K})^2}$ be the Frobenius norm of the tensor. The $k$-mode product of $\Mcal$ with a matrix $U_k \in \mathbb{R}^{I_k \times J}$, $J \leq I_k$, is denoted by $\Mcal \times_k U_k$ and is of size $I_1 \times \ldots \times I_{k-1} \times J \times I_{k+1} \times \ldots I_{K}$. Elementwise, $(\Mcal \times_k U_k)_{i_1\ldots i_{k-1}ji_{k+1}\ldots i_K} = \sum_{i_k=1}^{I_k} x_{i_1\ldots i_{k-1} i_{k} i_{k+1} \ldots i_K}u_{ji_k}$ for $j = 1, 2, \ldots, J$.  We refer to slices as two-dimensional sections of a tensor. For a three-way tensor $\Mcal \in \mathbb{R}^{I_1 \times I_2 \times I_3}$, we let $M_i \in \mathbb{R}^{I_1 \times I_2}$ denote the $i$th frontal slice, $i = 1, 2, \ldots, I_3$. 

\section{Methodology}\label{sec:methods}

\subsection{Tucker Decomposition}\label{sec:problemsetup} 
For a three-way tensor $\Mcal \in \mathbb{R}^{a \times b \times n}$ representing ABPM data, we let $a$ be the length of the time grid, $b$ be the number of measurements, and $n$ be the number of patients. In our motivating study, $a=24$ for 24 hours, $b=3$ corresponding to matched SBP, DBP, and HR, and $n=207$ for 207 OSA patients (Section~\ref{sec:data}). Our goal is to leverage the tensor structure to learn a low-dimensional representation of the data, for which we propose to take advantage of Tucker decomposition \citep{tucker_mathematical_1966}.

For tensor $\Mcal \in \mathbb{R}^{I_1 \times I_2 \times \ldots \times I_K}$, the Tucker decomposition approximates it via a core tensor $\Gcal \in \mathbb{R}^{r_1 \times r_2 \times \ldots \times r_K}$ and orthogonal matrices $U_k \in \mathbb{R}^{I_k \times r_k}$ for each mode $(k= 1, 2, \ldots, K)$. The best approximation for given ranks can be computed using the higher-order orthogonal iteration (HOOI) algorithm for the following problem \citep{de_lathauwer_best_2000}:
\begin{equation} \label{eq:hooi}
    \minimize_{\substack{\Gcal, U_k \\ U_k^\top U_k=I}} \|\Mcal - \Gcal \times_1  U_1 \times_2 U_2 \ldots \times_K U_K\|_F^2.
\end{equation} 

\begin{figure}[!t]
\centering
\includegraphics[width=0.7\textwidth]{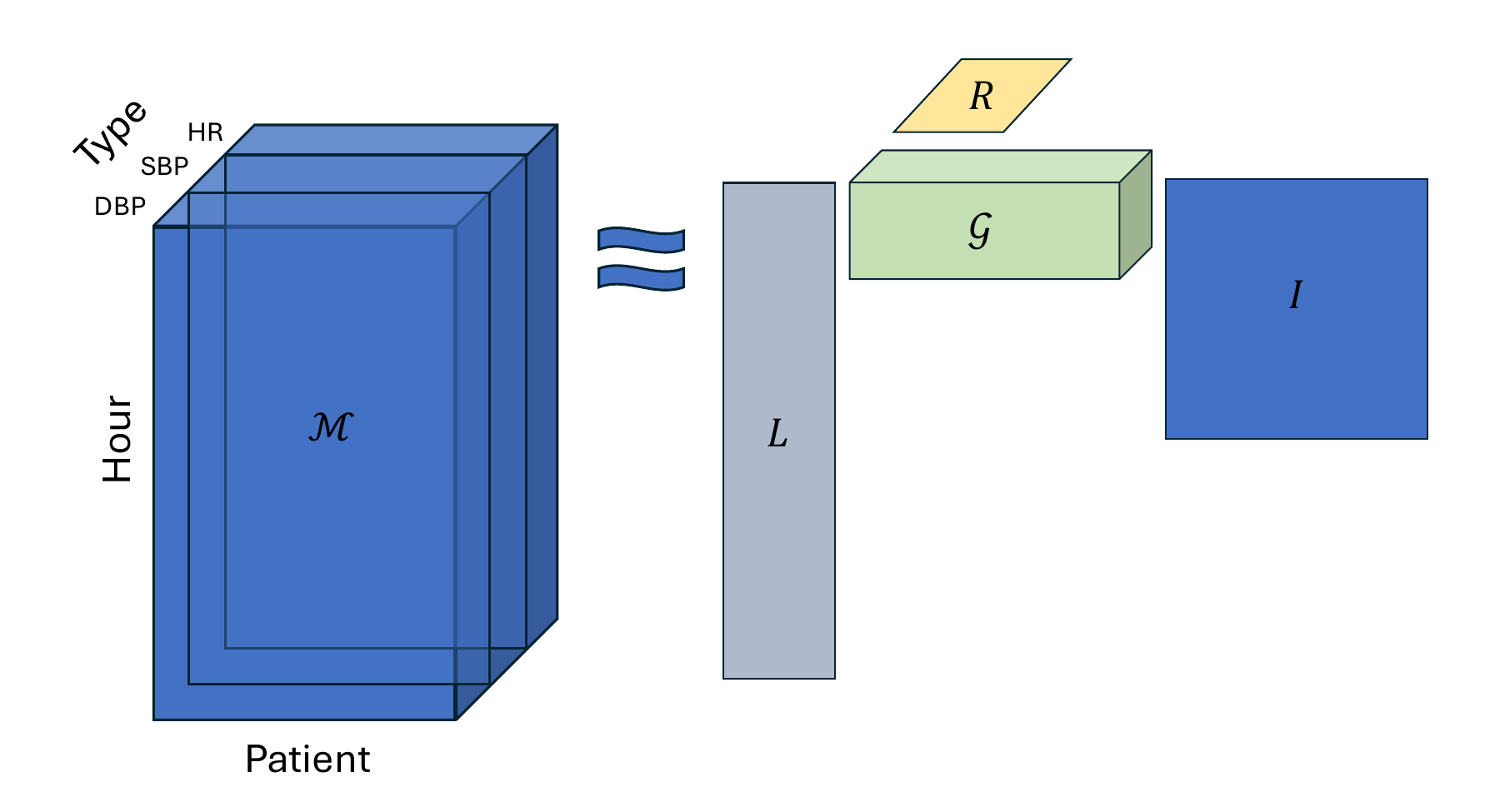}
\caption{Tensor representation of ABPM data with the low-rank approximation, where $L$ captures temporal components, $R$ represents measurement components, and $I$ denotes the identity matrix. The core tensor $\Gcal$ contains individual-level scores for all subjects along the temporal and measurement directions.}%: multiple characteristics/views are collected on the same samples.} 
\label{fig:GLRAM}
\end{figure}

For the ABPM tensor $\Mcal \in \mathbb{R}^{a \times b \times n}$, the matrix along the first mode ($U_1$, time) can be viewed as compressing temporal signal, whereas the matrix along the second mode ($U_2$, measurements) can be viewed as compressing along the measurement types. As we would like to use the compressed representation in downstream analyses on the subject level, we keep the subject mode uncompressed ($U_3 = I$). To aid differentiation between time and measurement compressions, we let $L = U_1 \in \mathbb{R}^{a \times r_1}$ and $R = U_2 \in \mathbb{R}^{b \times r_2}$, in which case problem~\eqref{eq:hooi} can be alternatively represented as
\begin{equation}\label{eq:glram}
    \minimize_{\substack{\Gcal, L, R\\ L^{\top}L=I,\  R^{\top}R=I}} \sum_{i=1}^n \|M_i - LG_iR^{\top}\|_F^2,
\end{equation}
where $M_i \in \mathbb{R}^{a \times b}$, $G_i \in \mathbb{R}^{r_1 \times r_2}$ for $i=1, \ldots, n$, and $r_1$ and $r_2$ are called the decomposition ranks. This formulation is also known as generalized low-rank matrix decomposition (GLRAM) \citep{ye_generalized_2005}, and Figure~\ref{fig:GLRAM} illustrates the decomposition. 

There are several roadblocks to the adoption of~\eqref{eq:glram} to ABPM data: (i) the compression does not account for time ordering, that is, the underlying smoothness assumption on measurements over time; (ii) formulation~\eqref{eq:glram} and corresponding optimization algorithm is for complete data, but not all hours are always observed in real data, leading to block-missing patterns in the tensor; (iii) the Tucker decomposition is non-unique: for any orthogonal matrices \( U_L \in \mathbb{R}^{r_1 \times r_1} \) and \( U_R \in \mathbb{R}^{r_2 \times r_2} \), we have
\begin{equation}\label{eq:invariance}
    LG_iR^{\top} = LU_LU_L^{\top}G_iU_RU_R^{\top}R^{\top} = \widetilde{L}\widetilde{G}_i\widetilde{R}^{\top}, \ \text{with} \ \widetilde{L} = LU_L,\; \widetilde{R} = RU_R,\; \widetilde{G}_i = U_L^{\top} G_i U_R.
\end{equation}
While this indeterminacy is inconsequential when decomposition is used for compression, it is relevant in our setting, where we intend to interpret \( L \) and \( R \), and use \( \mathcal{G} \) for downstream inference.

\subsection{Proposed Decomposition}

To address the limitations of~\eqref{eq:glram}, we (i) introduce the smoothness penalty that accounts for the temporal structure of the data, and (ii) adjust the objective function to accommodate missing values. We return to identifiability issue in Section~\ref{sec:ident}.

In decomposition~\eqref{eq:glram}, the goal is to obtain a low-rank approximation $\widehat \Mcal$ to the observed tensor $\Mcal$, where each frontal slice is approximated as $\widehat M_i = LG_iR^{\top}$. To encourage a smooth temporal fit, we introduce a penalty that promotes smoothness of the reconstructed signal $\widehat M_i$ across the time dimension, separately for each measurement type. Specifically, we penalize second-order differences, which approximate the second derivative of the signal over time.

For the case of $a=24$ hourly time points, we define the second-difference matrix $D$ as 
\begin{equation}\label{eq:SecDiffMat}
    D = \begin{pmatrix}
    2 & -1 & 0 & 0 & \cdots & -1 \\
    -1 & 2 & -1 & 0 & \cdots & 0 \\
   % 0 & -1 & 2 & -1 & \cdots & 0 \\
    \vdots & \vdots & \vdots & \vdots & \ddots & \vdots \\
    -1 & 0 & 0 & \cdots & -1 & 2\\
    \end{pmatrix} \in \mathbb{R}^{24 \times 24}.
\end{equation}

To formally incorporate this smoothness constraint, we augment the objective function in~\eqref{eq:glram} with a quadratic penalty term $\lambda \|D\widehat M_i\|_F^2 = \lambda \|DLG_iR^{\top}\|_F^2$, where $\lambda \geq 0$ is a tuning parameter that controls the strength of the penalization. This term discourages abrupt changes over time in the reconstructed signal, thereby promoting smooth trajectories. Although smoothing $L$ directly may seem more intuitive and would simplify optimization, in \ref{sec:smooth_on_L} we show that such an approach leads to poor fit and instability due to orthogonality constraints.

To accommodate missing values, we let $\Omega = \{(i_1, i_2, i_3): m_{i_1 i_2 i_3}\mbox{ is observed}\}$ denote the index set of all observed elements in tensor $\Mcal$. For each frontal slice $M_i$, we denote the corresponding subset of observed elements by $\Omega_i$. We then restrict the reconstruction error in~\eqref{eq:glram} to be only evaluated with respect to observed elements.

In summary, we formulate a smooth tensor decomposition as the solution to the following optimization problem:
\begin{equation} \label{eq:obj_func}
    \minimize_{\substack{\Gcal, L, R\\ L^{\top}L=I,\  R^{\top}R=I}} \sum_{i=1}^n\Big\{\|M_i - LG_iR^{\top}\|_{\Omega_i}^2 + \lambda \|DLG_iR^{\top}\|_F^2\Big\}.
\end{equation}
We further derive an efficient iterative algorithm to solve this problem, develop an automatic data-driven procedure to select decomposition ranks and the smoothing parameter $\lambda$, and introduce a post-processing rotation step to ensure identifiability. Due to connections with HOOI minimization problem and algorithm, we call our approach SmoothHOOI.

\subsection{Optimization Algorithm}\label{sec: alg}
We first derive an algorithm to solve~\eqref{eq:obj_func} under the assumption of complete data, and subsequently extend it to accommodate missing values.

Under the assumption of complete data, we first show that an optimal $G_i$ has a closed form in terms of optimal $L$ and $R$. Full proofs of all results are provided in \ref{sec:three_proof}.

\begin{proposition}\label{p:Gsolution}
Let $L\in\R^{a\times r_1}$, $R\in \R^{b \times r_2}$, $\{G_i\}^n_{i=1}$ be the optimal solution to~\eqref{eq:obj_func} with $\Omega$ being a complete index set of $\Mcal$. Then $G_i = (I + \lambda L^\top D^\top DL )^{-1}L^\top M_i R$ for $i=1, \dots, n$. 
\end{proposition}

We substitute the closed form of $G_i$ into~\eqref{eq:obj_func}, leading to equivalent formulation
\begin{align} 
 \maximize_{\substack{L, R\\ L^{\top}L=I,\  R^{\top}R=I}} \sum_{i=1}^n \Big\{ \Tr(L(I+\lambda L^\top D^\top DL)^{-1}L^\top M_i RR^\top M_i^\top)\Big\}. \label{eq:obj_func_replaceG}
\end{align}
Thus, we reduced~\eqref{eq:obj_func} to a maximization problem over only two parameters: $L$ and $R$.  We next show that by reparametrization of $L$, we have closed-form solutions with respect to each of the parameters when the other is fixed.

\begin{proposition} \label{p:LR_UR}
Let $A=I + \lambda D^\top D\in \R^{a \times a}$. Problem~\eqref{eq:obj_func_replaceG} is equivalent to 
\begin{align} 
 \maximize_{\substack{U, R\\ U^{\top}U=I,\  R^{\top}R=I}} \sum_{i=1}^n \Big\{ \Tr(UU^{\top} A^{-1/2}M_i RR^\top M_i^\top A^{-1/2})\Big\} \label{eq:obj_func_replaceL},
\end{align}
where $U\in \R^{a \times r_1}$, and the column space of $L$ is equal to the column space of $A^{-\frac{1}{2}}U$.
\end{proposition}

We can use the new parametrization to derive closed-form updates for $U$ and $R$.

\begin{proposition} \label{p:LR_UR2}
  Consider maximization problem~\eqref{eq:obj_func_replaceL} from Proposition~\ref{p:LR_UR}. 

    (1) For a given U, the optimal R consists of the top $r_2$ eigenvectors of 
    \begin{equation*}
         M_R =\sum_{i=1}^n M_i^\top A^{-\frac{1}{2}} UU^\top A^{-\frac{1}{2}} M_i;
    \end{equation*}

    (2) For a given R, the optimal $U$ consists of the top $r_1$ eigenvectors of
    \begin{equation*}
        M_U = \sum_{i=1}^n A^{-\frac{1}{2}}M_iRR^\top M_i^\top A^{-\frac{1}{2}}.
    \end{equation*}
    
\end{proposition} 

Combining Propositions~\ref{p:Gsolution}--\ref{p:LR_UR2} yields an iterative algorithm for solving~\eqref{eq:obj_func}, which is summarized in Algorithm~\ref{a:myglram}. Given $L$, we compute $U$, and then update $R$. Given $R$, we update $U$ and reconstruct $L$. Although the reconstruction of $L$ is only column-space equivalent, this poses no issue due to the rotational invariance of Tucker decomposition, as shown in~\eqref{eq:invariance}. We return to this invariance in Section~\ref{sec:ident}. The iterations continue until convergence, defined as the change in the objective function values~\eqref{eq:obj_func} between successive iterations falling below a threshold $\eta$. Since each update is in closed form, the objective function values are guaranteed to be non-increasing. Together with the fact that the objective is bounded below, this ensures convergence to a partial optimum.

\begin{algorithm}[!t]
\DontPrintSemicolon
\KwInput{Tensor $\Mcal \in \mathbb{R}^{a \times b \times n}$ with frontal slices $\{M_i\}_{i=1}^n \in \mathbb{R}^{a \times b}$, ranks $r_k$, k = 1,2; smoothing parameter $\lambda\geq 0$; difference matrix $D \in \mathbb{R}^{a \times a}$; maximum number of iterations $t_{max}$; convergence threshold $\eta > 0$}

  $L^{(0)} \gets [I_{r_1}, 0]^\top$\tcp*{Generate an initial $L^{(0)}$, if not given}

  $A \gets I + \lambda D^\top D$, 
  $U^{(0)} \gets$ SVD: $A^{\frac{1}{2}}L^{(0)} = U^{(0)}D^{(0)}V^{(0)\top}$ \tcp*{Reparametrization}

  % \tcc{Iterative Procedure}
    $t \gets 1$\;
  \While{$t \leq  t_{max}$ \text{and not convergent}}{
  
     $M_R^{(t)} \gets \sum_{i=1}^n M_i^{\top} A^{-\frac{1}{2}}U^{(t-1)}{U^{(t-1)}}^{\top} A^{-\frac{1}{2}}M_i $\;
     $R^{(t)} \gets$ the $r_2$ eigenvectors of $M_R^{(t)}$ for the largest $r_2$ eigenvalues \tcp*{Update of $R$}
      
    $M_U^{(t)} \gets \sum_{i=1}^n A^{-\frac{1}{2}}M_iR^{(t)}{R^{(t)}}^{\top}M_i^\top A^{-\frac{1}{2}}$\;
    $U^{(t)} \gets$ the $r_1$ eigenvectors of $M_U^{(t)}$ for the largest $r_1$ eigenvalues  \; 
    $L^{(t)} \gets$ the top $r_1$ eigenvectors of $A^{-\frac{1}{2}}U^{(t)}{U^{(t)}}^{\top}A^{-\frac{1}{2}}$  \tcp*{Update of $L$}

    \For{$i$ from $1$ to $n$}{
        $G_{i}^{(t)} \gets (I+\lambda {L^{(t)}}^{\top}D^\top DL^{(t)})^{-1}{L^{(t)}}^{\top} M_iR^{(t)}$\;
    }

    $t \gets t+1$ \;
}
$ L \leftarrow L^{ (t-1) } $,
$ R \leftarrow R^{ (t-1) } $, $ {\{ G_i \}}^{n}_{i=1} \leftarrow {\{ G_i^{(t-1)} \}}^{n}_{i=1}$ \;

      \KwOutput{$L$, $R$, and $\{G_i\}_{i=1}^n$}
\caption{Optimization algorithm for~\eqref{eq:obj_func} for complete data}
\label{a:myglram}
\end{algorithm}

To accommodate missing data, we adopt an iterative imputation approach \citep{ilin_practical_2010, zheng2018regularized}, embedding the complete-data algorithm as an inner loop. We begin by initializing the missing entries in the tensor using a constant value, such as zero. Given this initial imputation, we solve the optimization problem as in the complete-data case. Once convergence is reached within the inner loop, we update the initial imputation as follows
\begin{equation*} \label{eq:impute}
    M_i = H_i \circ M_i + (1-H_i) \circ LG_i R^{\top},
    % M_i^{(k)} = H_i \circ M_i + (1-H_i) \circ L^{(k)}G_i^{(k)}R^{(k)^{\top}}
\end{equation*}
where $H_i$ is the indicator matrix for $M_i$, with entries corresponding to observed $\Omega_i$ elements being equal to one, and zero otherwise. 

\subsection{Tuning Parameter and Rank Selection} \label{sec:tuning_approach}
The resulting decomposition from~\eqref{eq:obj_func} depends on the ranks $r_1$, $r_2$, and tuning parameter $\lambda$, which together control the tradeoff between fit to the data and smoothness. We propose to select these parameters via $k$-fold cross-validation, where the observed tensor elements are randomly split into $k$ mutually exclusive folds: each fold is used once for validation while the remaining $k-1$ folds form the training data on which the decomposition is fit. The optimal parameter values minimize the cross-validation error
\begin{equation*}
    CV(r_1,r_2,\lambda)=\frac{1}{k} \sum_{j=1}^k \Big\{\sum_{i=1}^n \|M_i - L^{(-j)}G_i^{(-j)}{R^{\top}}^{(-j)}\|^2_{\mathcal{B}_{ij}} \Big\},
\end{equation*}
where $\mathcal{B}_{ij}$ is the set of indices $(x,y)$ for the validation data in matrix $M_i$ for the $j$th fold, and $L^{(-j)}$, $G_i^{(-j)}$, and $R^{(-j)}$ are estimated using all data except the $j$th fold.

To reduce the computational cost, we adopt a coarse-to-fine grid search strategy \citep{bischl_hyperparameter_2023}, first exploring a broad range before refining the search. We also incorporate warm starts \citep{chu_warm_2015, ash_warm-starting_2020}: for nearby $\lambda$ values, we initialize the algorithm with the imputed data and $L$ from the previous solution to accelerate convergence.

Additional rank parsimony can be achieved by evaluating the proportion of explained variability for each component in the final decomposition (Section~\ref{sec:ident}) and discarding components with negligible contribution. In this case, the optimal smoothing parameter should be re-evaluated to account for the trade-off between ranks and smoothing.  

\subsection{Identifiability} \label{sec:ident}
The Tucker decomposition is non-unique, as shown in~\eqref{eq:invariance}. To ensure identifiability of $L$, $R$, and $\Gcal$ for downstream analyses, we break the rotational invariance by fixing the matrices $U_L$ and $U_R$ to (i) promote orthogonality between estimated scores in each core $G_i$ and (ii) order the scores in $G_i$ by decreasing percentage of explained signal. 

To achieve this, we utilize tensor unfoldings and SVD. Specifically, we apply SVD to the mode-1 and mode-2 unfoldings of the core tensor $\Gcal$, denoted $G_{(1)} \in R^{r_1 \times r_2n}$ and $G_{(2)} \in R^{r_2 \times r_1n}$, respectively. Applying SVD
$
    G_{(k)} = U_k D_k V_k^\top, k = 1,2,
$
produces orthogonal matrices $U_k$ whose columns are ordered by decreasing singular values, which are identifiable. Thus, using $U_L = U_1$ and $U_R = U_2$ leads to
\begin{equation*}
    \widetilde{L} = LU_L, \; \widetilde{R} = RU_R, \; \widetilde{G}_i = U_L^\top G_i U_R
\end{equation*}
such that the scores in $\widetilde G_i$ are orthogonal and ordered by the signal strength with respect to each mode. These rotations can be viewed as obtaining MLSVD decomposition from Tucker \citep{de_lathauwer_multilinear_2000}. The resulting
$\widetilde{L}$, $\widetilde{R}$, and $\widetilde{G}_i$ are thus identifiable (by conditions of SVD as long as singular values are distinct) and can be used in downstream analyses.

\section{Simulations}\label{sec:simu}

We simulate data to mimic the ABPM dataset described in Section~\ref{sec:data}, using an additive signal plus noise model with $a=24$, $b = 3$ and $n \in \{50, 200, 500\}$:
$$ 
\mathcal{M}_{noise}^{a \times b \times n} = \mathcal{M}_{smooth}^{a \times b \times n} + \mathcal{E}^{a \times b \times n}.
$$ 
 Each signal slice $M_i\in \R^{24 \times 3}$ in $\mathcal{M}_{smooth}$ is generated as
$
M_i = LG_iR^\top,
$
with fixed matrices $L\in \R^{24 \times r_1}$ and $R\in 
R^{3 \times r_2}$ obtained from SmoothHOOI applied to real data (Section~\ref{sec:data}) with $r_1 = 3$, $r_2 = 2$ and $\lambda = 4$. This construction yields a low rank tensor structure with smoothness in the temporal direction. The subject-specific core matrices $G_i$ for $i=1,\ldots,n$ are generated by sampling $(g_{11}, g_{12}, g_{21}, g_{22}, g_{31}, g_{32})$ i.i.d from $ \mathcal{N}(\widehat{\mathbf{\mu}}, \widehat{\Sigma})$, where $\widehat \mu$ and $\widehat \Sigma$ are the empirical mean and covariance estimated from the estimated core tensor $\widehat{ \mathcal{G}}$. To generate the noise tensor $\mathcal{E}^{24 \times 3 \times n}$, we draw its elements i.i.d from $N(0, \sigma^2)$, where $\sigma^2 \in \{\widehat \sigma^2_e, 0.1\widehat \sigma^2_e, 2\widehat \sigma^2_e\}$, and $\widehat \sigma^2_e$ denotes the empirical variance of the residuals between $\mathcal{M}$ and its reconstruction $\widehat{\mathcal{M}}$. We denote these settings as noise levels equal to 0.1, 1, and 2. Since the real data were centered and scaled before analysis, the simulated data are also on a normalized scale.

To generate final simulated tensor $\mathcal{M}_{sim}$, we apply two missing mechanisms to $\mathcal{M}_{noise}$. First, we randomly remove 0\%/10\%/20\%/50\% of the entries. Second, we introduce structured missingness by jointly removing (or retaining) SBP, DBP, and HR at a random subset of time points for each subject (allowing up to 20 of the 24 points to be missing). This leads to an average overall missing rate of 42\% (ranging from 37\% to 49\% across replications).

We define a baseline setting as $n=200$, noise level equal to 1, and 20\%  missing rate. Each simulation factor (sample size, noise level, and missing rate) is varied independently from this baseline to isolate its effect on performance, with 100 replications per combination.

We evaluate all methods based on two loss functions, overall reconstruction error $\Lcal_\Mcal$ and error in recovery of temporal components $\Lcal_L$, where
\begin{align*}
     \mathcal{L}_{\mathcal{M}} = \frac{1}{|\mathcal{M}_{smooth}|}\sum_{i=1}^n\|M_{smooth,i} - \widehat{M}_i\|_F^2, \quad\text{and} \quad
     \mathcal{L}_L=\frac{1}{\sqrt2}\|LL^\top - \widehat{L}\widehat{L}^\top\|_F.
\end{align*}
The loss $\Lcal_\Mcal$ follows the form of the mean squared error, while $\Lcal_L$ is chordal distance, 
which is rotation-invariant to enable fair comparison across methods. 

We assess the performance of our proposed SmoothHOOI~\eqref{eq:obj_func} with all parameters selected by $k$-fold cross-validation as in Section~\ref{sec:tuning_approach} with $\lambda\in [1, 50]$. We also consider its oracle version, where all parameters are selected to minimize $\mathcal{L}_{\mathcal{M}}$. As a benchmark that does not account for tensor structure, we consider FPCA \citep{shang_survey_2014}, which we implement using the \textsf{fpca.sc()} function of the \textsf{refund} R package \citep{goldsmith_refund_2024}.

While $\mathcal{L}_{\mathcal{M}}$ is invariant to the choice of ranks,  the temporal loss $\Lcal_L$ is only meaningful when the estimated rank $\widehat r_1$ matches the true rank $r_1$. Thus, we consider two cases:
\begin{itemize}
    \item Case 1 (fixed ranks): All methods use true values $r_1=3$ and $r_2=2$. %and vary $\lambda \in [1, 50]$. 
    \item Case 2 (flexible ranks): The ranks are allowed to vary within $r_1 \in [2,6]$ and $r_2 \in [2,3]$. %, and $\lambda \in [1, 50]$. 
\end{itemize} 

In Case 1, since FPCA is applied separately to each measurement type, it yields three separate $L$ matrices: $\widehat{L}_{SBP}$, $\widehat{L}_{DBP}$, and $\widehat{L}_{HR}$. To create a single matrix for comparison, we apply QR decomposition to each and average the resulting orthonormal bases to obtain a unified $\widehat{L}_{FPCA}$. 
In Case 2, 
we select the number of components for FPCA to explain at least 99\% of the variance (default setting). 
Since the number of selected components can differ across SBP, DBP, and HR, we report the minimum of the three as the effective rank $\widehat r_1$, which favors FPCA in rank estimation comparison.

\begin{figure}[!t]
    \centering
    \begin{subfigure}[b]{\textwidth}
    \refstepcounter{subfigure}
        \begin{tikzpicture}
            \node[anchor=north west, inner sep=2pt] at (0,0) {\includegraphics[width=\textwidth]{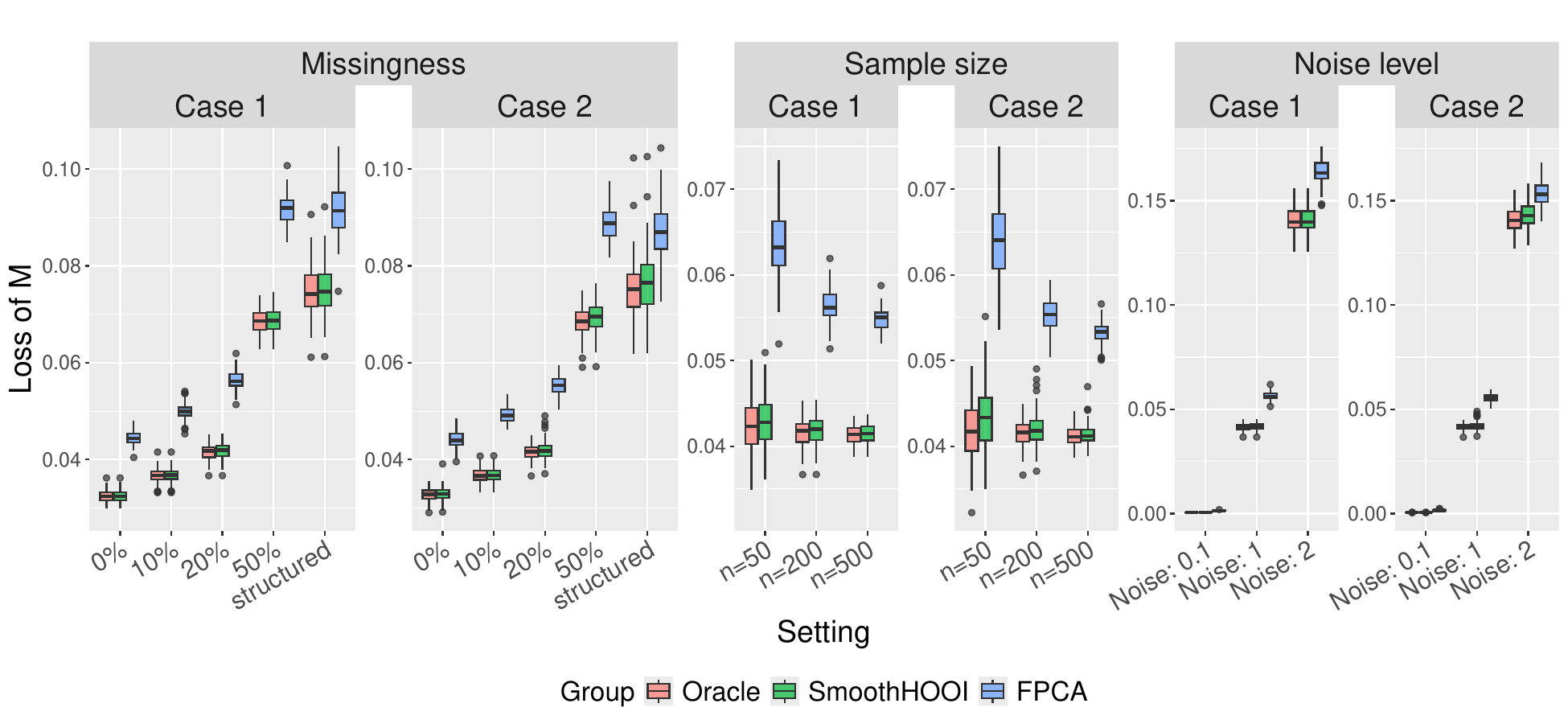}};
            \node[anchor=north west] at (0,0) {\textbf{(a)}};
        \end{tikzpicture}
        \vspace{-35pt}
        \caption*{}
        \label{fig:lossM}
    \end{subfigure}
    
    \begin{subfigure}[b]{0.8\textwidth}
    \refstepcounter{subfigure}
        \begin{tikzpicture}
            \node[anchor=north west, inner sep=2pt] at (0,0) {\includegraphics[width=\textwidth]{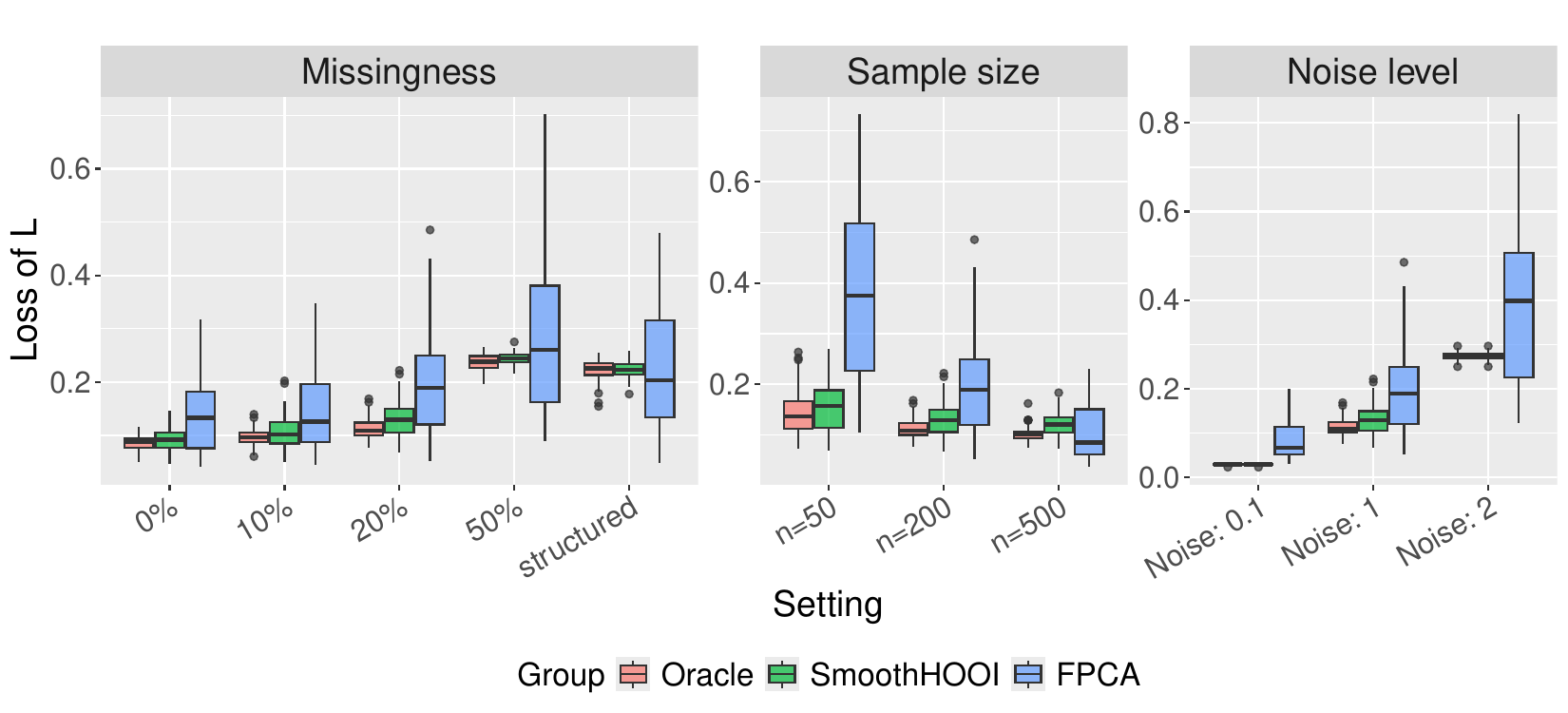}};
            \node[anchor=north west] at (0,0) {\textbf{(b)}};
        \end{tikzpicture}
        \caption*{}
        \label{fig:lossL}
    \end{subfigure}
    \vspace{-25pt}
    \caption{Boxplots for losses for $\Mcal$ (a) and $L$ (b) across 100 replications. Case 1 corresponds to fixed ranks, and Case 2 to flexible ranks. Loss of $\Lcal$ is only computed in Case 1. Oracle refers to SmoothHOOI with tuning parameters selected to minimize $\mathcal{L}_{\Mcal}$.}
    \label{fig:losses}
\end{figure}

Figure~\ref{fig:lossM} displays the reconstruction error $\Lcal_\Mcal$ in all settings across 100 replications. As expected, the errors increase as missing rate and noise level increase, and decrease with larger sample sizes. The structured missingness leads to similar median reconstruction error as random missingness at 50\%, but with a higher variability. Across all settings, $\widehat{\Mcal}$ from SmoothHOOI selected via 10-fold cross-validation closely matches the performance of the oracle version, supporting excellent performance of the proposed automatic tuning method. FPCA consistently yields higher reconstruction error than proposed SmoothHOOI. 

Figure~\ref{fig:lossL} displays the loss of $L$ in Case 1 across 100 replications. The loss from the cross-validated version of our method is slightly higher than the oracle. FPCA still demonstrates the worst performance in all cases except in the presence of structured missingness with large sample size, where the median $\mathcal{L}_L$ for FPCA is lower than both the oracle and cross-validated estimates. However, this comes at the cost of much higher variability, with some replications resulting in significantly higher errors than the proposed SmoothHOOI. \ref{sec:add_sim} shows that both SmoothHOOI and FPCA tend to slightly overestimate ranks with high missingness, but SmoothHOOI identifies the correct rank more often.

\section{Data Analysis}\label{sec:data}

\subsection{Data Description and Analysis Goals}
The Hyperglycemic Profiles in Obstructive Sleep Apnea (HYPNOS) study recruited adults with type 2 diabetes and moderate-to-severe OSA and randomized them into one of two groups, lifestyle counseling with positive airway pressure therapy and lifestyle counseling alone. A detailed description of the trial protocol can be found in \citet{rooney_rationale_2021}. Here we only focus on the baseline data prior to randomization. Ambulatory 24-hour blood pressure monitoring was performed with the device (Welch Allyn ABPM 6100) to measure SBP, DBP, and HR at hourly intervals over a period of at least 24 hours. OSA severity is represented by 4\% oxygen desaturation index (ODI4), which were measured based on a home sleep apnea test using Apnealink Plus monitor (Resmed, San Diego, CA). The ODI4 variable was right-skewed, and we applied a natural log transformation in the downstream regression analyses. Study participants also completed an actigraphy study using the Actiwatch (Philips Respironics, Murraysville, PA), and sleep periods were estimated using the actigraphy data and the proprietary algorithm of the Phillips Actiware software. The actigraphy-estimated sleep period were used for independent validation of tensor decomposition output.

Overall, there are 207 participants with ABPM data collected at baseline, with 191 of them having complete data on baseline clinical characteristics. 
Details of data preprocessing and descriptive statistics of the clinical characteristics are available in \ref{sec: add_data}. The final ABPM dataset is organized on a 24-hour grid, from 12 pm to 11am, and included 2365 missing values, representing 15.9\% of all data points (24 hours $\times$ 3 measurement types $\times$ 207 patients). Only 18.4\% of participants had complete data; among those with missingness, the proportion of missing values ranged from 4.2\% to 79.2\%.  

Our goals are to utilize baseline data to (i) characterize ABPM patterns in this population; (ii) examine associations between the ABPM measurements and the baseline characteristics such as age, sex, race, body mass index (BMI), and OSA severity, represented by ODI4. 

\subsection{Traditional Analysis Approach} \label{sec: trad_analysis}
We first applied a traditional analysis approach by summarizing ABPM data with mean values for each measurement type (SBP, DBP, HR) separately for daytime and nighttime periods, with nighttime defined as the period from 12am to 6am. This resulted in six summary measures per patient (three measurement types $\times$ two time periods), which we subsequently used as outcomes in linear regression models with age, sex, race, BMI and log-transformed ODI4 as covariates. Models were fit on 191 participants with complete covariate data. Table~\ref{tab:mean_lr_res} shows estimated coefficients for each model. We evaluated statistical significance using both unadjusted and Bonferroni-corrected p-values (adjusting for six tests).  At the 5\% significance level, ODI4 was associated with DBP during both day and night, but these associations did not remain significant after correction for multiple comparisons.
However, ODI4 is clinically expected to be associated with elevated blood pressure \citep{silverberg2002treating, kawano_influence_2010, kwon_obstructive_2024}, suggesting that the traditional summary-based approach may lack the statistical power.

\subsection{Analysis Using Proposed SmoothHOOI}

We next considered the proposed SmoothHOOI, with low-rank scores in the estimated core tensor $\mathcal{G}$ taken as outcomes in downstream regression. Before applying the algorithm, the data for the three measurements, DBP, SBP, and HR, were normalized using their respective sample means and standard deviations across all 207 patients with available ABPM measurements at baseline. This normalization ensured that a single tuning parameter $\lambda$ could be applied uniformly across DBP, SBP, and HR. The resulting data formed a three-way tensor with dimensions $24 \times 3 \times 207$ ($\text{hour} \times \text{measurement type} \times \text{patient}$).

We first applied five-fold cross-validation from Section~\ref{sec:tuning_approach}, which yielded $r_1 =6$, $r_2=3$, and $\lambda=12$. Figure~\ref{fig:exp_var} shows the percentage of variability in the data explained by each resulting component in $L$ and $R$, respectively, after applying the identifiability correction of Section~\ref{sec:ident}. Given that the last components explained little variability and that the cross-validation may overestimate the ranks in the presence of missing data (Section~\ref{sec:simu}), we opted for a more parsimonious model with $r_1=3$, $r_2 = 2$ and corresponding optimal $\lambda = 4$. This final reduced model explained 59.0\% of total variation in the data, calculated by $\sum_i \|\widehat{M_i}\|_{\Omega_i}^2/\sum_i \|M_i\|_{\Omega_i}^2$, leading to six scores for each subject's core matrix $G_i$. These scores can be viewed as the degree of interaction between the temporal components (from $L$) and the physiological components (from $R$) when reconstructing the subject-specific $M_i$ as
\begin{equation} \label{eq: reconstruct}
    M_i = \underbrace{g_{11,i}L_1R_1^\top + g_{12,i}L_1R_2^\top +
    g_{21,i}L_2R_1^\top + g_{22,i}L_2R_2^\top +
    g_{31,i}L_3R_1^\top + g_{32,i}L_3R_2^\top}_{\widehat M_i} + E_i. 
\end{equation}

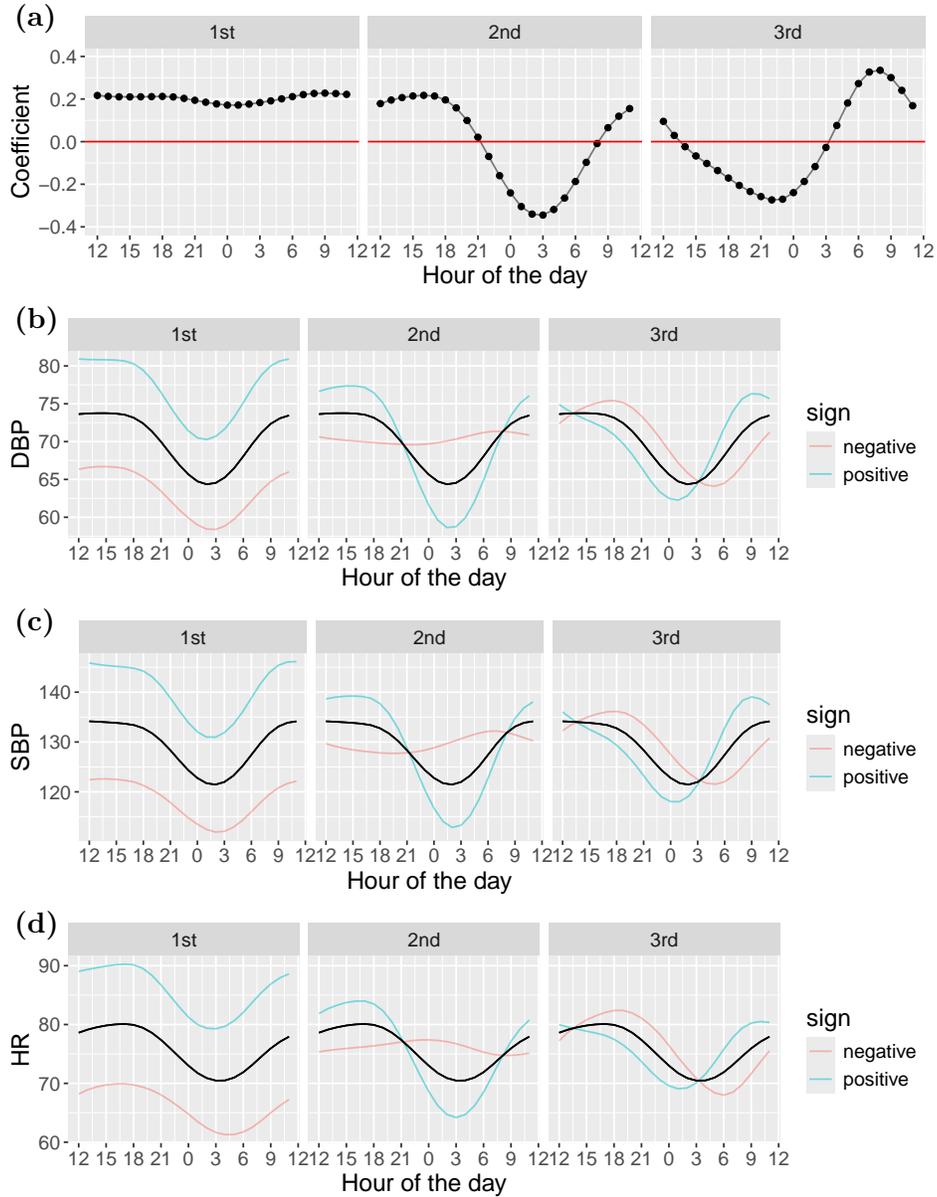
\begin{figure}[!t]
    \centering
    \begin{subfigure}[b]{0.75\textwidth}
    \refstepcounter{subfigure}
        \begin{tikzpicture}
            \node[anchor=north west, inner sep=2pt] at (-0.1,-0.1) {\includegraphics[width=\textwidth]{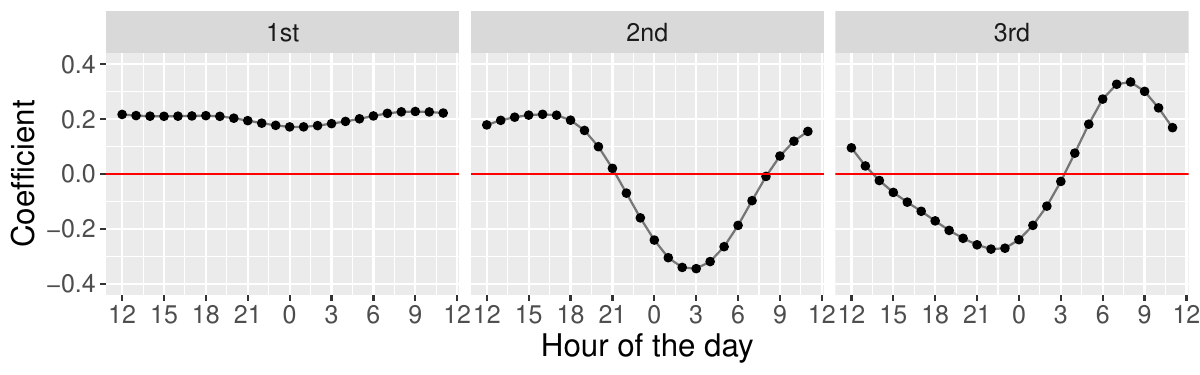}};
            \node[anchor=north west] at (0,0) {\textbf{(a)}};
        \end{tikzpicture}
         \vspace{-35pt}
        \caption*{}
        \label{fig:L_comp}
    \end{subfigure}
    \begin{subfigure}[b]{0.75\textwidth}
    \refstepcounter{subfigure}
        \begin{tikzpicture}
            \node[anchor=north west, inner sep=2pt] at (-0.1,-0.1) {\includegraphics[width=\textwidth]{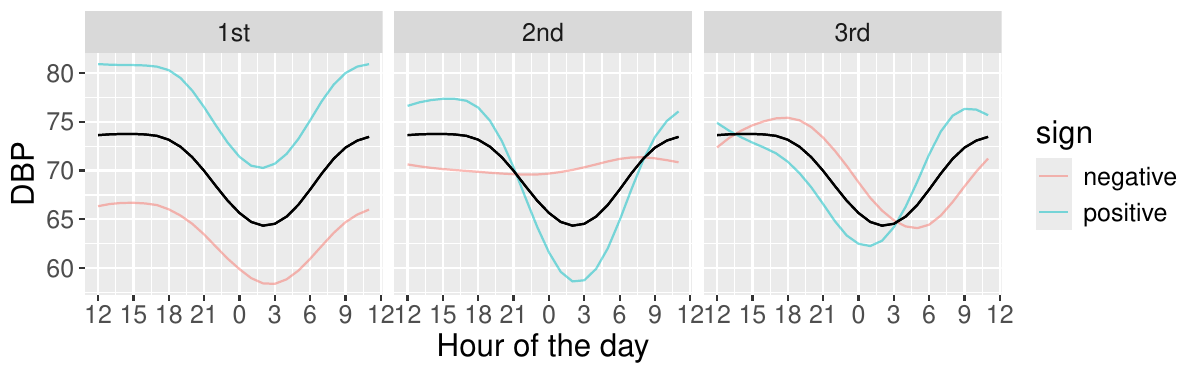}};
            \node[anchor=north west] at (0,0) {\textbf{(b)}};
        \end{tikzpicture}
        \vspace{-35pt}
        \caption*{}
        \label{fig:L_comp_DBP}
    \end{subfigure}
    \begin{subfigure}[b]{0.75\textwidth}
    \refstepcounter{subfigure}
        \begin{tikzpicture}
            \node[anchor=north west, inner sep=2pt] at (-0.1,-0.1) {\includegraphics[width=\textwidth]{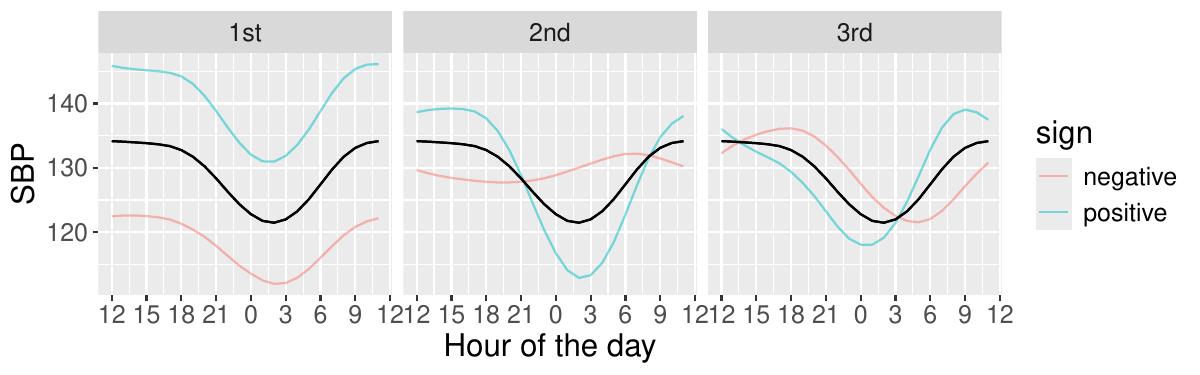}};
            \node[anchor=north west] at (0,0) {\textbf{(c)}};
        \end{tikzpicture}
        \vspace{-35pt}
        \caption*{}
        \label{fig:L_comp_SBP}
    \end{subfigure}
    \begin{subfigure}[b]{0.75\textwidth}
    \refstepcounter{subfigure}
        \begin{tikzpicture}
            \node[anchor=north west, inner sep=2pt] at (-0.1,-0.1) {\includegraphics[width=\textwidth]{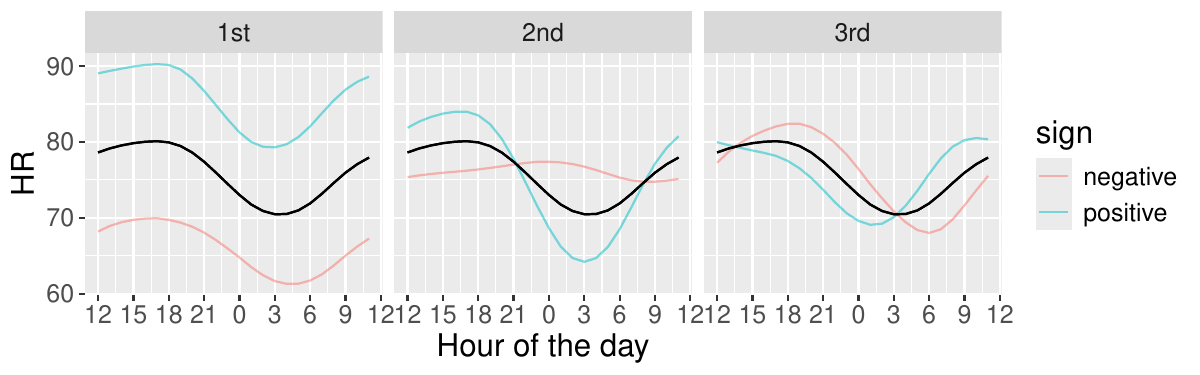}};
            \node[anchor=north west] at (0,0) {\textbf{(d)}};
        \end{tikzpicture}
        \caption*{}
        \label{fig:L_comp_HR}
    \end{subfigure}
    
    \vspace{-25pt}
    \caption{(a): Coefficients for each estimated time component $L_k$ from baseline HYPNOS ABPM data, $k=1, 2, 3$. (b)--(d): Effect of temporal components $L_k$ on DBP (b), SBP (c), and HR (d). The solid black line represents the sample mean of DBP, SBP, and HR over the 24-hour period with colored lines representing positive (blue curve) and negative (red curve) shifts with respect to $L$ components.}
    \label{fig:L_comp_figs}
\end{figure}

Figure~\ref{fig:L_comp} shows the estimated three temporal components of $L$. To aid interpretation, we evaluate how variation along each component influences the sample mean of the reconstructed signals, $\sum_{i}\widehat M_i/n$, for each measurement type based on~\eqref{eq: reconstruct}. For example, to interpret $L_1$,  we isolate the part of the signal affected by $L_1$, given by the combination $(g_{11, i}, R_1^{\top} + g_{12, i}R_2^{\top})$, and compute empirical standard deviations across subjects for each measurement. We then shift the sample mean up and down by one standard deviation in this direction.  We repeat this process for the other components. Finally, we rescale and center to return back to the original measurement scale. Figure~\ref{fig:L_comp_DBP}--\ref{fig:L_comp_HR} illustrate how each temporal component in $L$ modulates the average DBP, SBP, and HR trajectories. The first component reflects the overall blood pressure/heart rate levels as it leads to a uniform shift in the mean up or down. The second captures nocturnal dipping behavior: positive shifts in the direction of $L_2$ lower nighttime values and raise daytime values, while negative shifts yield a flatter, non-dipping profile. The third component changes the dipping time to earlier or later, suggesting that it accounts for differences in individual sleep times and chronotypes.

The two estimated components for $R$ represent (i) a weighted average of the three physiological measures (37/100 DBP + 33/100 SBP + 3/10 HR); (ii) a contrast between blood pressure and heart rate (-13/100 DBP - 33/100 SBP + 27/50 HR).

\begin{figure}[!t]
    \centering
    \includegraphics[width=0.6\linewidth]{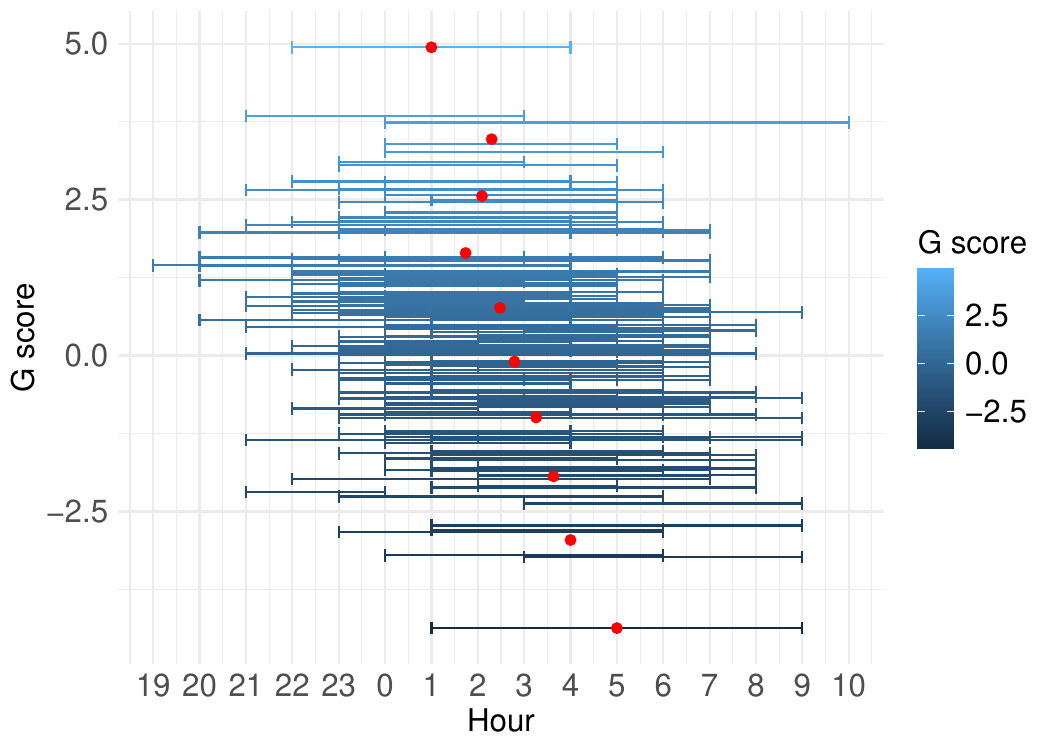}
    \caption{Relationship between $g_{31}$ scores and actigraphy-estimated sleep periods. Blue bars represent the time of sleep periods for each subject. Red dots denote the average midpoint of sleep for each of 10 equal-width intervals spanning the range of $g_{31}$ scores (e.g., $[-4.37, -3.44]$, $[-3.44, -2.51]$, etc.).}
    \label{fig:chrono}
\end{figure}

To validate the interpretation of the third temporal component, we examine the association between $g_{31}$ scores, which reflect concurrent time shifts across all measurements, and actigraphy-estimated sleep periods. Figure~\ref{fig:chrono} shows $g_{31}$ scores against corresponding individual sleep periods. Subjects with higher $g_{31}$ scores tended to sleep earlier, which aligns with the leftward shift induced by a positive change in the $L_3$ component (Figure~\ref{fig:L_comp_DBP}--\ref{fig:L_comp_HR}), confirming that the third component effectively extracts variability due to differences in individual sleep times.

We used linear regression models to assess associations between the extracted scores in $G_i$ (restricting to the first and second temporal components, as the third reflects sleep timing) and clinical characteristics. The $g_{11}$ scores reflect concordant changes in overall levels across all three measurements, while $g_{12}$ scores reflect the overall contrast between blood pressure and heart rate. Similarly, $g_{21}$ scores represent concordant nocturnal dipping magnitude across all three measurements, and $g_{22}$ scores capture dipping heterogeneity between blood pressure and heart rate. No covariates were significant for the nocturnal dipping scores ($g_{21}$ and $g_{22}$, Table~\ref{tab:g2_table}), so we further only focus on $g_{11}$ and $g_{12}$.

\begin{figure}[!t]
    \centering
    \begin{subfigure}[b]{0.8\textwidth}
    \refstepcounter{subfigure}
        \begin{tikzpicture}
            \node[anchor=north west, inner sep=2pt] at (-0.1,-0.1) {\includegraphics[width=\textwidth]{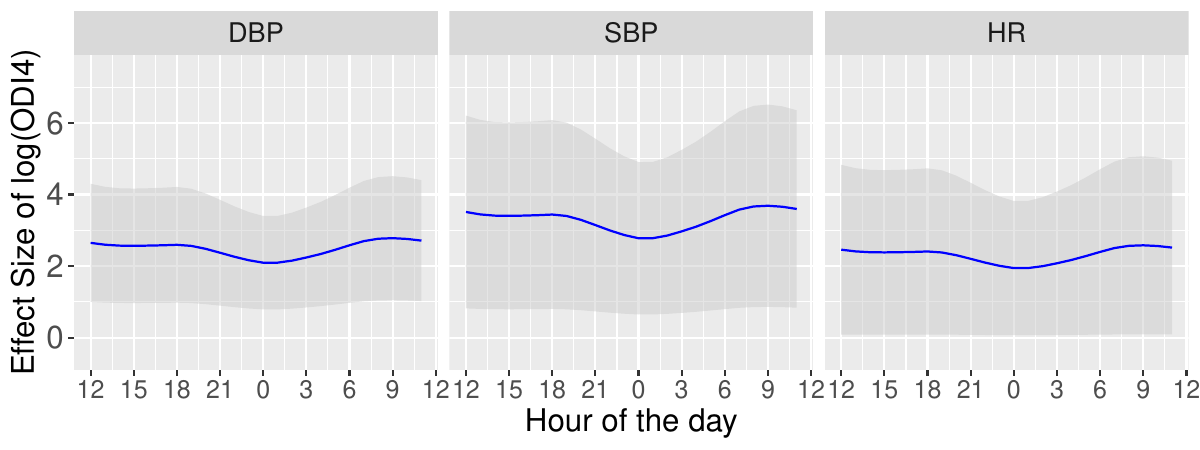}};
            \node[anchor=north west] at (0,0) {\textbf{(a)}};
        \end{tikzpicture}
         \vspace{-35pt}
        \caption*{}
        \label{fig:odi_effect}
    \end{subfigure}
    \begin{subfigure}[b]{0.8\textwidth}
    \refstepcounter{subfigure}
        \begin{tikzpicture}
            \node[anchor=north west, inner sep=2pt] at (-0.1,-0.1) {\includegraphics[width=\textwidth]{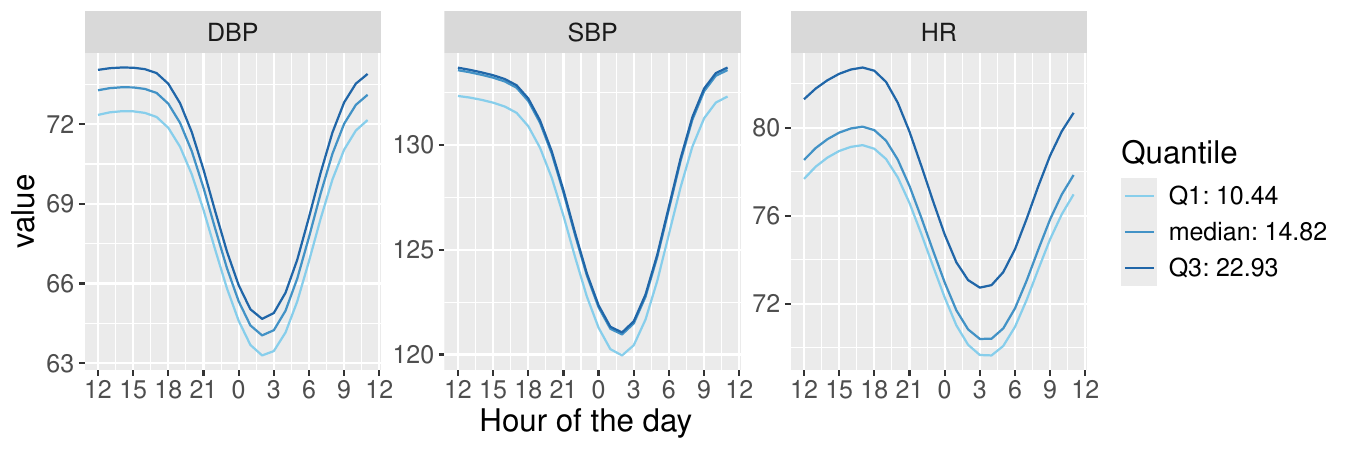}};
            \node[anchor=north west] at (0,0) {\textbf{(b)}};
        \end{tikzpicture}
        \caption*{}
        \label{fig:odi_change}
    \end{subfigure}
    \vspace{-25pt}
    \caption{(a): The estimated effect size of log-transformed ODI4 on DBP, SBP, and HR; (b): Estimated DBP, SBP, and HR profiles when ODI4 is set to its first quartile (Q1), median, and third quartile (Q3) values (unit: events/hour), with all other covariates held at their means.}
    \label{fig:effect_size_odi}
\end{figure}

Table~\ref{tab:lr_res} summarizes the regression results for $g_{11}$ and $g_{12}$. Unlike the findings in Section~\ref{sec: trad_analysis}, higher ODI4 was significantly associated with elevated average of blood pressure and heart rate, even after Bonferroni adjustment. Figure~\ref{fig:effect_size_odi} illustrates the effect size of log-transformed ODI4 on the original measurement scale for each DBP, SBP and HR with corresponding 95\% confidence intervals, as well as predicted temporal trends evaluated at quartiles. The observed positive association between OSA severity and the increases in DBP, SBP, and HR is in line with what is expected from clinical literature \citep{silverberg2002treating, kawano_influence_2010, kwon_obstructive_2024}.

\begin{table}[!t]
    % \begin{adjustwidth}{}{}
    \centering
    \begin{tabular}{lcccc}
    \toprule
    & \multicolumn{2}{c} {\textbf{Model $g_{11}$}} & \multicolumn{2}{c} {\textbf{Model $g_{12}$}}  \\
    & \textbf{$\boldsymbol{\widehat{\beta}^{(1)}}$} & \textbf{95\% CI of $\boldsymbol{\widehat{\beta}^{(1)}}$} & \textbf{$\boldsymbol{\widehat{\beta}^{(2)}}$} & \textbf{95\% CI of $\boldsymbol{\widehat{\beta}^{(2)}}$}  \\
    \midrule
    \textbf{Age} & $-0.057$ & $(-0.120, 0.006)$ &  $-0.110^{***,}\text{\textasciicircum\textasciicircum\textasciicircum}$ & $(-0.164, -0.056)$  \\
    \textbf{Sex} & & \\
    Male & Ref. & Ref. & Ref. & Ref. \\
    Female & $-0.275$ & $(-1.486, 0.936)$ & $1.383^{**,}\text{\textasciicircum\textasciicircum}$ & $(0.345, 2.423)$  \\
    \textbf{Race} & & \\
    White & Ref. & Ref. & Ref. & Ref. \\
    Others & $1.948^{**,}\text{\textasciicircum\textasciicircum}$ & $(0.784, 3.111)$ & $-0.505$ & $(-1.503, 0.493)$  \\
    \textbf{BMI} & $0.106$ & $(-0.013, 0.226)$ & $-0.124^{*,}\text{\textasciicircum}$ & $(-0.227, -0.021)$ \\ 
    \textbf{log(ODI4)} & $1.615^{**,}\text{\textasciicircum\textasciicircum}$ & $(0.652, 2.578)$ & $0.060$ & $(-0.767, 0.886)$ \\ 
    \bottomrule
    \vspace{10pt}
    \end{tabular}
    \caption{Coefficient estimates from the linear regression models with the 95\% confidence intervals (CIs). Model $g_{11}$: The outcome is the overall joint blood pressure/heart rate effect (37/100 DBP + 33/100 SBP + 3/10 HR). Model $g_{12}$: The outcome is the overall contrast between blood pressure and heart rate (-13/100 DBP - 33/100 SBP + 27/50 HR). The significance level before Bonferroni adjustments: *$p<0.05$, **$p<0.01$, ***$p<0.001$. The significance level after Bonferroni adjustments: \textasciicircum$p<0.05$, \textasciicircum\textasciicircum$p<0.01$, \textasciicircum\textasciicircum\textasciicircum$p<0.001$.}
    \label{tab:lr_res}
    % \end{adjustwidth}
\end{table}

From Table~\ref{tab:lr_res}, race was additionally significant for $g_{11}$, whereas age, sex and BMI for $g_{12}$. Figures \ref{fig:effect_plot_others}-\ref{fig:change_plot_others} illustrate effect sizes and predicted temporal trends for these covariates, and here we provide an overall interpretation. Non-white individuals (primarily Black in our study) had a significantly higher DBP, SBP, and HR compared to White individuals, consistent with prior findings \citep{golbus_wearable_2021}. Older age was significantly associated with lower HR, consistent with previous studies \citep{tanaka_age-predicted_2001,santos_does_2013}, while the association between age and overall blood pressure was weaker, likely due to opposing trends in DBP and SBP. DBP tends to rise until midlife and then decline after age 50 \citep{whelton_epidemiology_1994}, a pattern we observed with a downward trend beginning around age 53. Meanwhile, SBP generally increases with age \citep{whelton_epidemiology_1994}, leading to a muted net effect on average BP.
Females had lower blood pressure and higher heart rate than males, consistent with established sex-based physiological differences \citep{maranon_sex_2013, prabhavathi_role_2014}. Although sex associations were not statistically significant for any single measurement in our dataset, modeling across all measurements using the $R$ components allowed us to borrow strength and detect a significant difference in the contrast ($g_{12}$). Higher BMI was associated with significant increases in both DBP and SBP, while its effect on HR was weaker and negative. Although this may seem counterintuitive, these are adjusted rather than marginal effects. Additionally, prior work suggests that BMI primarily affects the upper tail of the heart rate distribution \citep{ghosal_distributional_2025}, which may not be well captured at the hourly resolution of ABPM data.

\section{Discussion}
\label{sec:discus}

% Put your final comments here. 
We propose a novel low-rank tensor decomposition approach, SmoothHOOI, which incorporates temporal smoothing along one of the tensor modes and can handle missing data. 
While motivated by ABPM data, the method is applicable to other settings involving temporally structured matched multivariate data. The corresponding R package \textsf{SmoothHOOI} is available at \href{https://github.com/IrinaStatsLab/SmoothHOOI}{https://github.com/IrinaStatsLab/SmoothHOOI}.

In our application, treating ABPM data as a tensor allows us to borrow information across both the temporal and measurement dimensions. This results in a more parsimonious and interpretable representation of the data compared to traditional summaries, while also improving statistical power in downstream analysis. SmoothHOOI eliminates the need to impose a uniform division of daytime and nighttime across all individuals, and can effectively disentangle shifts in sleep times from other temporal components. It also separates differences in overall blood pressure levels from differences in nocturnal dipping magnitude.
A key example demonstrating the method's sensitivity is the detection of association between OSA severity and elevated blood pressure and heart rate, which is expected based on epidemiological evidence \citep{silverberg2002treating, kawano_influence_2010, kwon_obstructive_2024}, but is not identified using summary statistic-based approaches. Finally, the simulation studies confirm that borrowing strength across measurement types improves estimation accuracy compared to FPCA approach applied to one measurement at a time.

There are several directions for future work. First, while our presentation focused on a three-way tensor structure motivated by the application, the methodology and algorithm easily extend to higher-order tensors with smoothing along a single temporal mode. Second, while we used hourly resolution due to the frequency of ABPM measurements, the method can be applied to higher-resolution data without modification. Since our approach accommodates missing data, time alignment across subjects is not required. While higher resolution would increase computational cost, it is feasible since the analysis of hourly resolution data in Section~\ref{sec:data} only took a few minutes on a standard laptop without parallelization. The main computational bottleneck lies in the automatic selection of tuning parameters. Generalized cross-validation (GCV) \citep{ruppert_semiparametric_2003} could offer a faster alternative to $k$-fold cross-validation. In our experience, it performed well in fully observed settings, but was less reliable with missing data, suggesting a need for further methodological development.

\section*{Acknowledgements}
This research was supported by R01HL172785.

% \vspace*{-8pt}

%\section*{Funding}

% \section*{Conflict of interest}
% None declared.

\bibliographystyle{chicago}
\bibliography{ABPM}

\clearpage

\appendix

\renewcommand\thesection{Appendix \Alph{section}}
\renewcommand\thesubsection{\Alph{section}.\arabic{subsection}}
\renewcommand{\thetable}{S\arabic{table}}  
\setcounter{table}{0}
\renewcommand{\thefigure}{S\arabic{figure}}
\setcounter{figure}{0}
\renewcommand{\theequation}{\Alph{section}.\arabic{equation}}
\setcounter{equation}{0}

\begin{center}
\LARGE{
Supporting information for ``Smooth tensor decomposition with application to ambulatory blood pressure monitoring data''
}
\end{center}
\bigskip
\begin{abstract}
    \noindent \ref{sec:smooth_on_L} provides empirical studies illustrating challenges with smoothing components directly. \ref{sec:three_proof} provides the proofs of Propositions 1, 2, and 3. \ref{sec:add_sim} reports the estimated ranks produced by SmoothHOOI and FPCA in Case 2 of the Simulation Studies. Section~\ref{sec: add_data} provides additional details on analysis of HYPNOS data.
\end{abstract}
\bigskip

\section{Smoothness constraints on L components} \label{sec:smooth_on_L}

In this section, we demonstrate empirically that enforcing both smoothness and orthogonality on components within Tucker decomposition leads to poor fit and instability, a conflict that to our knowledge has not been explicitly identified in prior work.

We consider the unpenalized Tucker decomposition formulation (2) and add a smoothing penalty based on second-order difference matrix $D$ in (4) directly on $L$, leading to optimization
problem
\begin{equation} \label{eq:obj_funcL}
    \minimize_{\substack{\Gcal, L, R\\ L^{\top}L=I,\  R^{\top}R=I}} \sum_{i=1}^n\Big\{\|M_i - LG_iR^{\top}\|_{\Omega_i}^2\Big\} + \lambda \|DL\|_F^2.
\end{equation}
Due to direct penalization of $L$, problem~\eqref{eq:obj_funcL} is easier than (5) and can be solved directly using the original HOOI algorithm~\citep{de_lathauwer_best_2000}.

We next investigate how the fitted model from~\eqref{eq:obj_funcL} changes over algorithm iterations when the smoothing parameter $\lambda$ is large using ABPM data
from Section 4. Specifically, we fix the ranks as $r_1 =6$, $r_2=3$, and use 
$\lambda=25000$. We then select three random subjects with missing data, and track how these values are imputed by the corresponding algorithm over iterations.

\begin{figure}[!t]
    \centering
    \includegraphics[width=\linewidth]{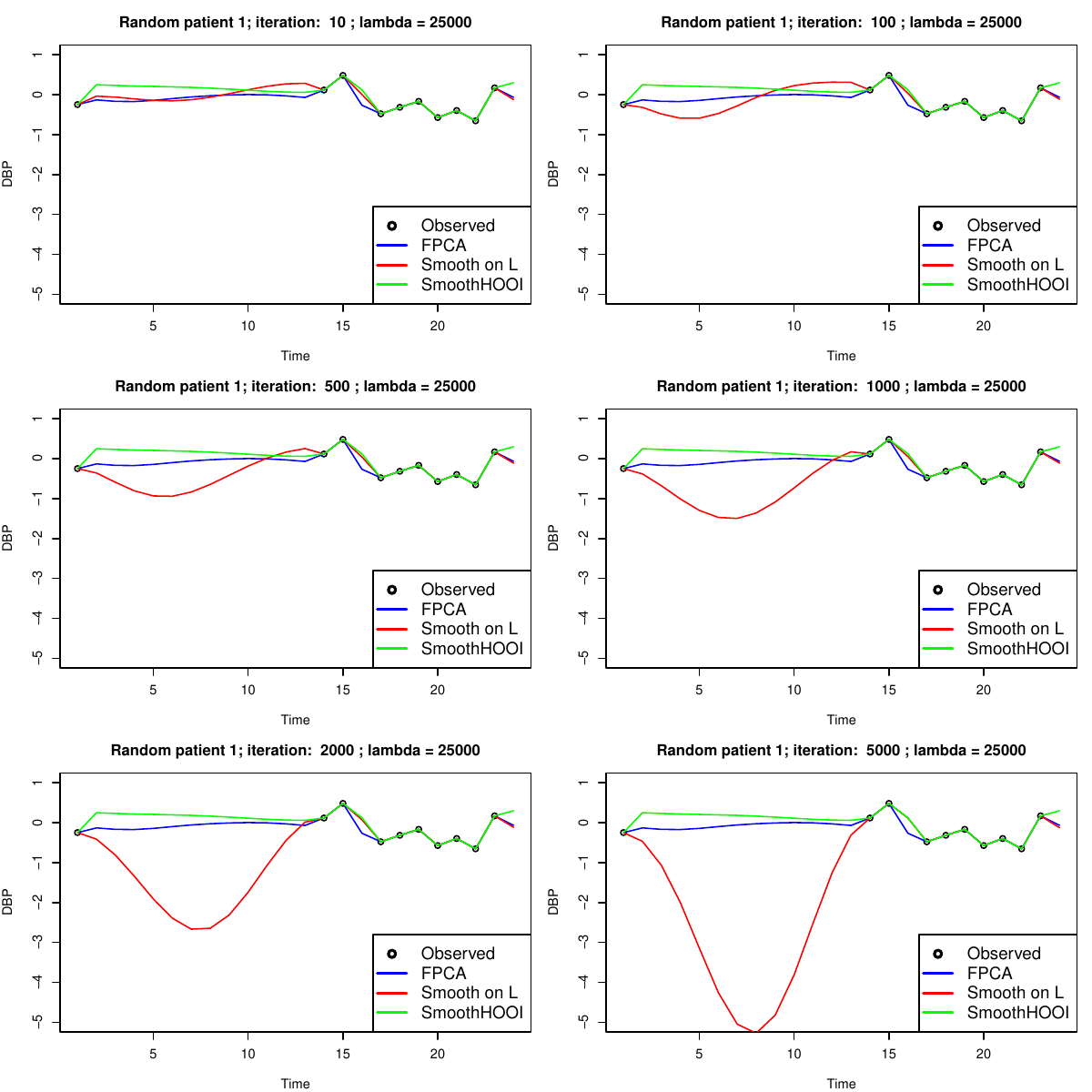} 
    \caption{Data imputation results with smoothness applied to $L$: Example 1. The figure shows the originally observed data and the imputed values at unobserved time points. Results from FPCA and SmoothHOOI are included for comparison.}
    \label{fig:example_1}
\end{figure}

\begin{figure}[!t]
    \centering
    \includegraphics[width=\linewidth]{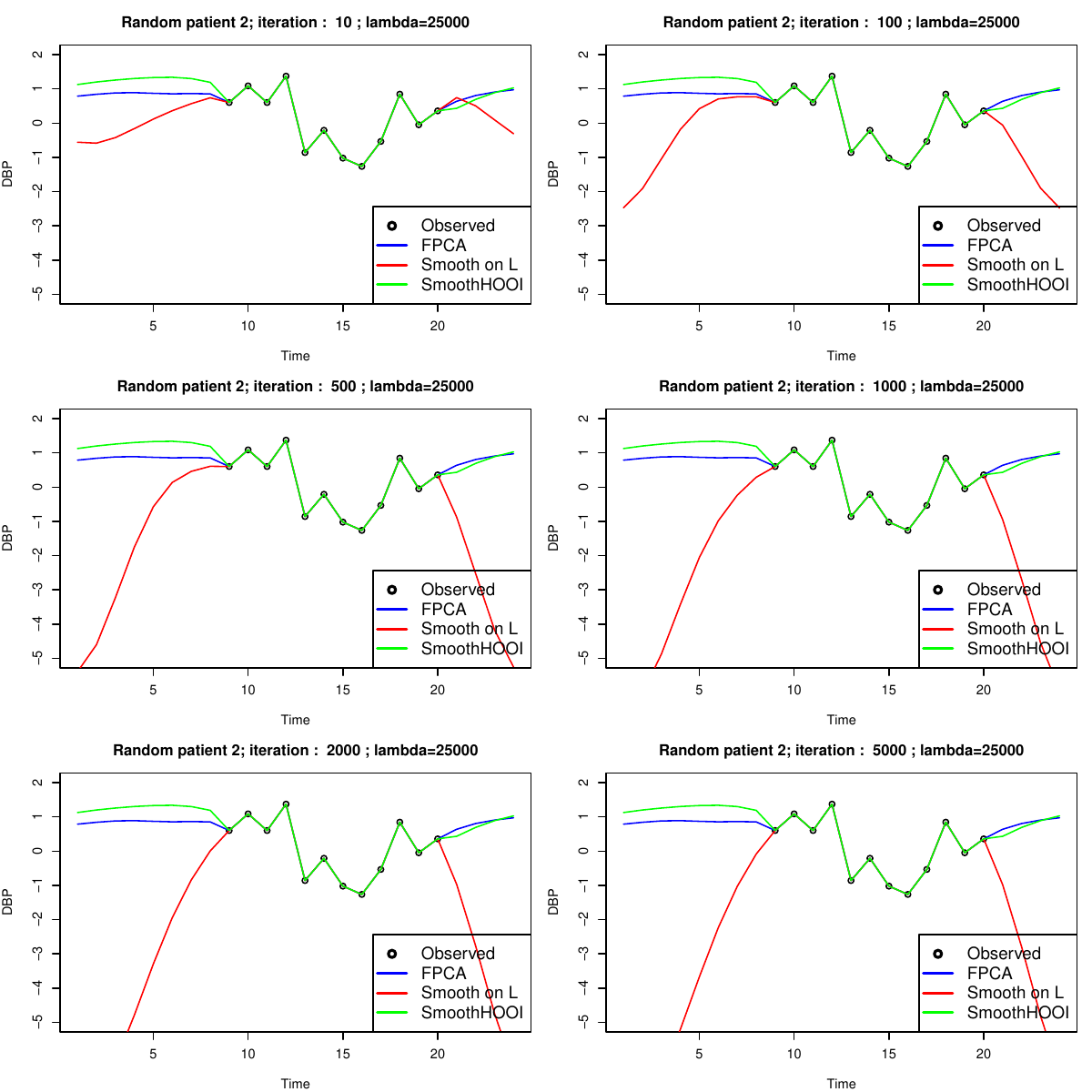} 
    \caption{Data imputation results with smoothness applied to $L$: Example 2. The figure shows the originally observed data and the imputed values at unobserved time points. Results from FPCA and SmoothHOOI are included for comparison.}
    \label{fig:example_2}
\end{figure}

\begin{figure}[!t]
    \centering
    \includegraphics[width=\linewidth]{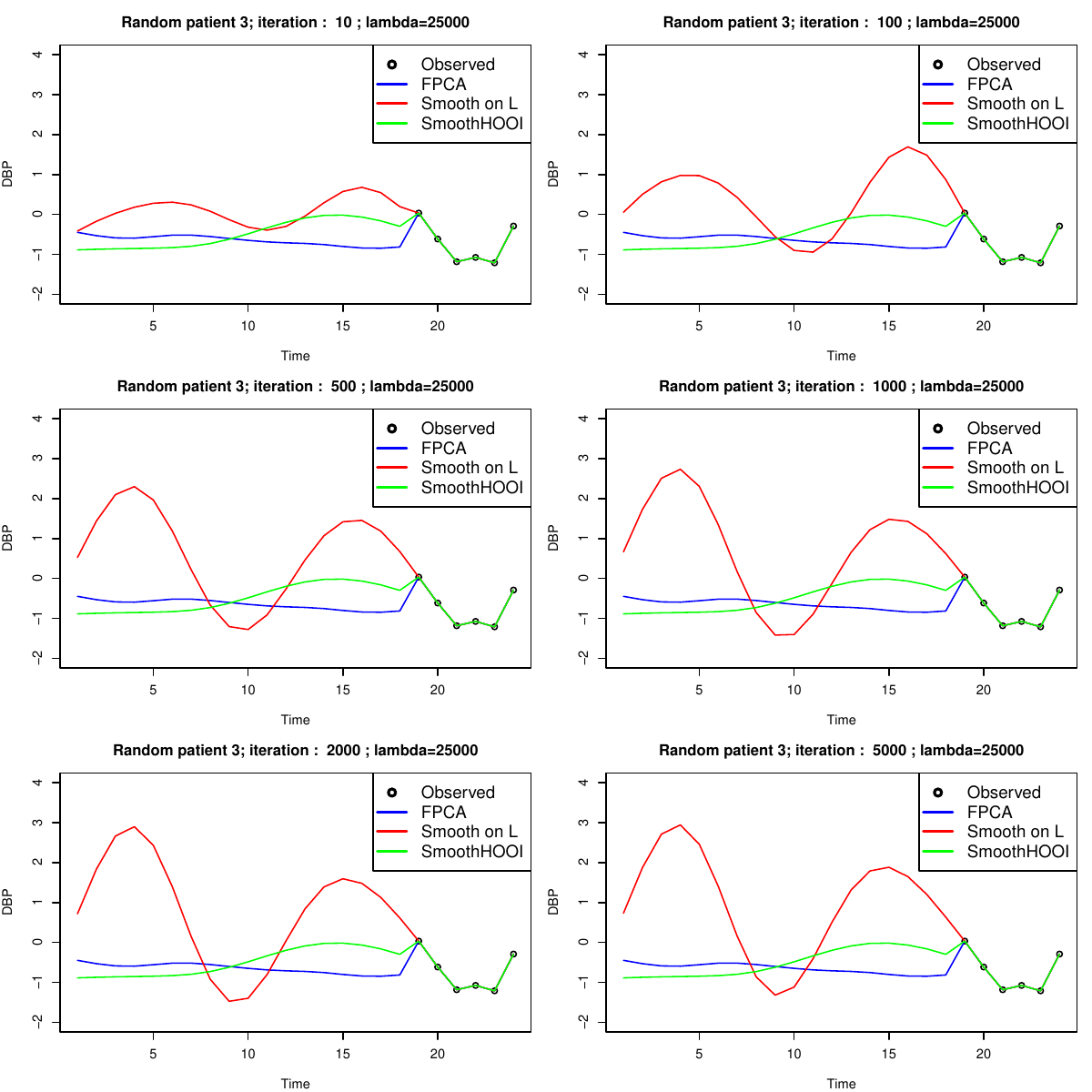} 
    \caption{Data imputation results with smoothness applied to $L$: Example 3. The figure shows the originally observed data and the imputed values at unobserved timepoints. Results from FPCA and SmoothHOOI are included for comparison.}
    \label{fig:example_3}
\end{figure}

Figures~\ref{fig:example_1}-\ref{fig:example_3} present the obtained imputed values for three randomly selected subjects with missing data over the 10th, 100th, 500th, 1000th, 2000th, and 5000th iterations of the algorithm, using DBP measurement. The observed values (dots) are left unchanged. For reference, we also display imputed values obtained by the proposed SmoothHOOI (smoothing on the fit rather than components), and non-tensor-based FPCA. Zero is used as the initial imputation starting point.

Intuitively, in the absence of data, we would expect smoothing to produce a flat fit—an effect seen in both FPCA and the proposed SmoothHOOI. However, when smoothing is applied directly to the components of $L$, the result is a sinusoidal pattern that becomes more pronounced with each iteration. This behavior, though initially counterintuitive, arises from the orthogonality constraint imposed on the columns of 
 $L$. In the extreme case where $L$ has rank one, the smoothed component is constant, yielding a flat fit. But when the rank exceeds one, orthogonality prevents subsequent components from also being constant. Instead, the combination of the orthogonality constraint and the smoothness penalty leads the columns of 
$L$ to align with the eigenvectors of $D^{\top}D$, which are known to be discrete sine and cosine functions. This explains the emergence of the sinusoidal structure.

This issue is resolved when smoothness is imposed on the fitted values rather than on the components themselves. This approach better aligns with the underlying assumption that the data—not the components—are smooth. The resulting $L$ is also smooth as a byproduct (Section 4), but the fit no longer exhibits the exaggerated sinusoidal behavior. 

\section{Proofs of Proposition 1, 2, and 3}\label{sec:three_proof}

In this section, we present the detailed proofs of the propositions. 

\begin{proof}[Proof of Proposition 1]
Using the properties of matrix trace,
\begin{equation*}
    \begin{split}
    & \|M_i - LG_iR^{\top}\|_F^2 + \lambda \|DLG_iR^{\top}\|_F^2 \\
    = & \Tr \{ (M_i - LG_iR^\top)(M_i - LG_iR^\top)^\top \} + \lambda \Tr \{ (DLG_iR^\top)(DLG_iR^\top)^\top \} \\
    = & \Tr(M_iM_i^\top) - 2\Tr(M_i^\top LG_iR^\top) + \Tr(LG_iR^\top RG_i^\top L^\top) + \lambda \Tr(DLG_iR^\top RG_i^\top L^\top D^\top) \\
    = &  \Tr(M_iM_i^\top) - 2\Tr(M_i^\top LG_iR^\top) + \Tr(G_iG_i^\top) + \lambda \Tr(L^\top D^\top DLG_iG_i^\top ) \\
    = & \Tr(M_iM_i^\top) - 2\Tr(M_i^\top LG_iR^\top) +  \Tr \{ (I+\lambda L^\top D^\top DL)G_i G_i^\top \}.
    \end{split}
\end{equation*}
By calculating the gradient with regard to $G_i$ and setting it to 0, we have 
\begin{equation*}
    -2(R^\top M_i ^\top L) + 2 G_i^\top (I+\lambda L^\top D^\top DL) = 0 \quad \Rightarrow
\end{equation*}
\begin{equation*}
    (I+\lambda L^\top D^\top DL) G_i = L^\top M_i R \quad \Rightarrow
\end{equation*}
\begin{equation*}
    G_i  = (I+\lambda L^\top D^\top DL)^{-1} L^\top M_i R.
\end{equation*}
\end{proof}

\begin{proof}[Proof of Proposition 2]
First, we rewrite 
\begin{equation*}
    \sum_{i=1}^n  \Tr \Big\{ L(I+\lambda L^\top D^\top DL)^{-1}L^\top M_i RR^\top M_i^\top \Big\}
\end{equation*}
into
\begin{equation*}
\begin{split}
    % & \sum_{i=1}^n \Tr \Big\{ L(I+\lambda L^\top D^\top DL)^{-1}L^\top M_i RR^\top M_i^\top \Big\} \\
    % = 
    &  \sum_{i=1}^n  \Tr \Big\{ L(L^\top L +\lambda L^\top D^\top DL)^{-1}L^\top M_i RR^\top M_i^\top  \Big\} \\
    = & \sum_{i=1}^n \Tr \Big\{ L (L^\top(I +\lambda D^\top D)L)^{-1}L^\top M_i RR^\top M_i^\top \Big\}.
\end{split}
\end{equation*}
Let $A=I + \lambda D^\top D$, then we have
\begin{equation*}
\begin{split}
    & \sum_{i=1}^n \Tr \Big\{ L (L^\top(I +\lambda D^\top D)L)^{-1}L^\top M_i RR^\top M_i^\top \Big\} \\
    =& \sum_{i=1}^n \Tr \Big\{ A^{\frac{1}{2}} L (L^\top A L)^{-1}L^\top A^{\frac{1}{2}} A^{-\frac{1}{2}} M_i RR^\top M_i^\top A^{-\frac{1}{2}} \Big\}.
\end{split}
\end{equation*}
Let $U$ be the left singular vector matrix from the SVD of $X=A^{\frac{1}{2}}L=UDV^\top$, then 
\begin{equation*}
    A^{\frac{1}{2}} L (L^\top A L)^{-1}L^\top A^{\frac{1}{2}} = X(X^\top X)^{-1}X^\top = UU^\top.
\end{equation*}
Finally, the maximization problem
\begin{equation*} 
 \maximize_{\substack{L, R\\ L^{\top}L=I,\  R^{\top}R=I}} 
 \sum_{i=1}^n  \Tr \Big\{ L(I+\lambda L^\top D^\top DL)^{-1}L^\top M_i RR^\top M_i^\top \Big\}
\end{equation*}
becomes 
\begin{equation}
    \maximize_{\substack{U, R\\ U^{\top}U=I,\  R^{\top}R=I}} \sum_{i=1}^n  \Tr (UU^{\top} A^{-1/2}M_i RR^\top M_i^\top A^{-1/2}).
    \label{eq:supp_eq_1}
\end{equation}
Let $P_{col(M)}$ denote the projection matrix onto the column space of $M$. Then given $A=I + \lambda D^\top D$ and $X=A^{\frac{1}{2}}L=UDV^\top$,
\begin{equation*}
    LL^\top = L(L^\top L)^{-1}L^\top = P_{col(L)} = A^{-\frac{1}{2}} X X^\top A^{-\frac{1}{2}} = A^{-\frac{1}{2}} U D^2 U^\top A^{-\frac{1}{2}}.
\end{equation*}
By the orthogonality constraint, $L^\top L = I$, which is equivalent to $X^\top A^{-1} X = I$, then we get 
\begin{equation*}
    V D U^\top A^{-1} U D V^\top = I.
\end{equation*}
Since $V^\top V = V V^\top = I$ and $D$ is diagonal, 
\begin{equation*}
    U^\top A^{-1} U = D^{-2}. 
\end{equation*}
Hence, 
\begin{equation*}
    P_{col(L)} = A^{-\frac{1}{2}} U (U^\top A^{-1} U)^{-1} U^\top A^{-\frac{1}{2}} = P_{col(A^{-1/2}U)}.
\end{equation*}

\end{proof}

\begin{proof}[Proof of Proposition 3]
By rearranging Equation \eqref{eq:supp_eq_1}, we get
\begin{equation*}
\begin{split}
    % & \Tr \left[ A^{\frac{1}{2}} L (L^\top A L)^{-1}L^\top A^{\frac{1}{2}} \sum_{i=1}^n A^{-\frac{1}{2}} M_i RR^\top M_i^\top A^{-\frac{1}{2}} \right] \\
     & \Tr \Big\{ UU^\top \sum_{i=1}^n A^{-\frac{1}{2}} M_i RR^\top M_i^\top A^{-\frac{1}{2}} \Big\} \\
    = & \Tr \Big\{ U^\top \Big( \sum_{i=1}^n A^{-\frac{1}{2}} M_i RR^\top M_i^\top A^{-\frac{1}{2}} \Big) U \Big\} \\
    = & \Tr \Big\{ R^{\top}\left( \sum_{i=1}^nM_i^{\top}A^{-1/2}UU^{\top}A^{-1/2}M_i\right)R \Big\}.
\end{split}
\end{equation*}
Then, this is optimization on the Stiefel manifold \citep{edelman_geometry_1998} and by Theorem 3.3 in \citet{ye_generalized_2005}, 

(1) for a given $U$, optimal $R$ consists of the top $r_2$ eigenvectors of 
    \begin{equation*}
         M_R =\sum_{i=1}^n M_i^\top A^{-\frac{1}{2}} UU^\top A^{-\frac{1}{2}} M_i,
    \end{equation*}
    
    % corresponding to the largest $r_2$ eigenvalues;
    
(2) for a given $R$, optimal $U$ consists of the top $r_1$ eigenvectors of 
    \begin{equation*}
        M_U = \sum_{i=1}^n A^{-\frac{1}{2}}M_iRR^\top M_i^\top A^{-\frac{1}{2}}.
    \end{equation*}.

\end{proof}

\section{Additional results on simulation studies}\label{sec:add_sim}

\begin{figure}[!t]
    \centering
    \includegraphics[width=\linewidth]{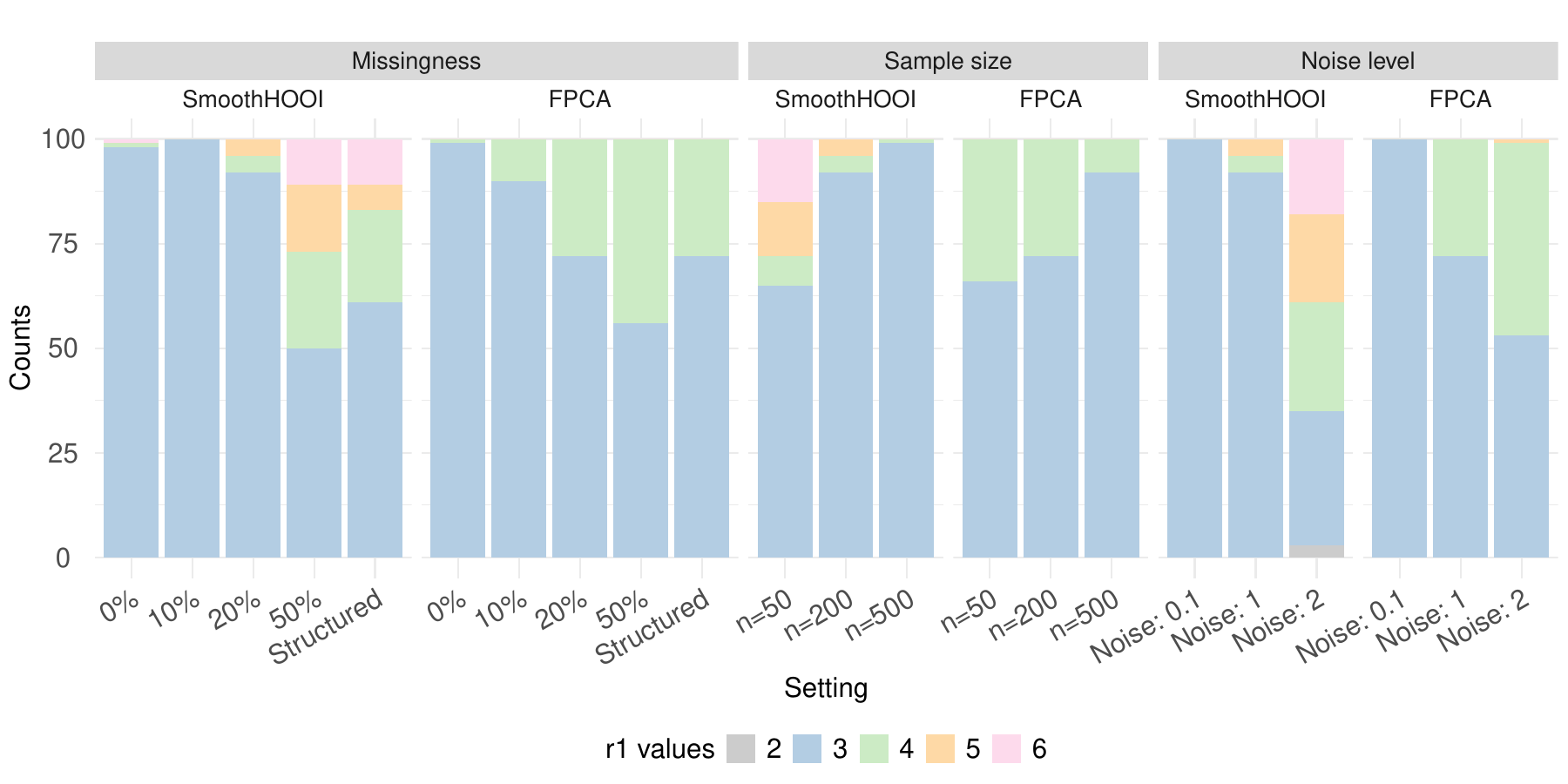} 
    \caption{The values of $r_1$ selected by SmoothHOOI and FPCA across 100 replicates in Case~2, the true value is $r_1 = 3$. For FPCA, the smallest rank across $b=3$ measurements is shown.}
    \label{fig:rankselect}
\end{figure}

Figure~\ref{fig:rankselect} displays the estimated rank $\widehat r_1$ in Case 2 across 100 replications. The results show that under challenging conditions—such as high missingness, small sample size, or high noise level—both methods tend to select larger values of $r_1$, likely as a compensatory response to increased uncertainty and reduced signal quality in the data. The proposed SmoothHOOI appears to yield higher estimated ranks than FPCA in these settings; however, the comparison is favorable to FPCA, as for the latter we report the smallest rank among the three separately estimated $L$ matrices (for SBP, DBP, and HR). Despite this, the proposed SmoothHOOI still achieves a higher overall proportion of correctly estimated ranks.

\section{Additional results on HYPNOS data analysis}\label{sec: add_data}

\subsection{ABPM Data Preprocessing}

We checked for data quality using criteria by \citet{casadei_24-hour_1988} and ABPM data processing pipeline in R package \textsf{bp} \citep{schwenck_bp_2022}. Specifically, the measurement was considered unreliable if SBP $>$ 240 or $<$ 50 mmHg; DBP $>$ 140 or $<$ 40 mmHg; HR $>$ 220 bpm or $<$ 27 bpm. There were only 0.3\% of such measurements in total, and they were set to missing. ABPM data were collected with the goal of covering a 24-hour period, with automatic readings scheduled hourly. However, some participants had shorter recordings due to missing data, while others contributed additional manual measurements or extended wear periods. To harmonize these discrepancies, the data were organized on a 24-hour grid. For participants with multiple measurements per hour or recordings exceeding 24 hours, hourly averages were computed across days for each corresponding clock hour and aligned to the start of each hour, from 12:00 p.m. to 11:00 a.m.
After accounting for non-wear periods and quality issues, the final dataset included 2365 missing values, representing 15.9\% of all data points (24 hours $\times$ 3 measurement types $\times$ 207 patients). %Only 18.4\% of participants had complete data; among those with missingness, the proportion of missing values ranged from 4.2\% to 79.2\%.
%Among the clinical characteristics, we found that ODI4 was highly right skewed, and we applied a natural log-transformation in the downstream regression analyses. 
Table~\ref{tab:desc} displays the descriptive statistics for the clinical characteristics, stratified by sex. 

\begin{table}[!t]
%\begin{adjustwidth}{}{}
\centering
\begin{tabular}{lccc}
\toprule
\multicolumn{2}{c}{\textbf{Characteristic}} & \textbf{Female}, N = 90 & \textbf{Male}, N = 101\\
\midrule
{\textbf{Age} (years)}
   & Median [Q1, Q3] & 61.0 [54.0, 67.0] & 61.0 [53.0, 67.0] \\
\multirow{2}{*}{\textbf{Race}} & White & 41 (46\%) & 62 (61\%)\\
 & Others & 49 (54\%) & 39 (39\%)\\
{\textbf{BMI} ($\mathrm{kg}/\mathrm{m}^2$)} 
   & Median [Q1, Q3] & 34.8 [31.1, 38.6] & 31.9 [22.9, 35.3] \\
{\textbf{ODI4} (events/hour)} 
   & Median [Q1, Q3] & 14.5 [10.5, 18.4] & 15.1 [10.1, 26.6] \\
\bottomrule
\vspace{10pt}
\end{tabular}
\caption{Descriptive statistics of age, race, BMI, and 4\% oxygen desaturation index (ODI4), stratified by sex for 191 of 207 participants from HYPNOS with complete covariate information. Q1: first quartile; Q3: third quartile.}
\label{tab:desc}
%\end{adjustwidth}
\end{table}

For validation against actigraphy-estimated sleep periods, we excluded 10 subjects who had multiple sleep and wake transitions within the same hour, 7 subjects who did not have estimated sleep times, and 5 subjects with multi-day actigraphy recordings.

\subsection{Regression model coefficients for the traditional summary statistic-based approach}

Table~\ref{tab:mean_lr_res} presents the coefficient estimates for the six regression models using mean daytime and nighttime SBP, DBP, and HR as outcomes.

\begin{table}[!t]
\begin{adjustwidth}{}{}
\centering
\begin{tabular}{lllllll}
\toprule
& \textbf{SBP Day} & \textbf{SBP Night} & \textbf{DBP Day} & \textbf{DBP Night} & \textbf{HR Day} & \textbf{HR Night} \\
\midrule
\textbf{Age} & $0.281^{**,}\text{\textasciicircum}$ & $0.362^{***,}\text{\textasciicircum\textasciicircum}$ & $-0.201^{**,}\text{\textasciicircum}$ & $-0.061$ & $-0.291^{***,}\text{\textasciicircum\textasciicircum}$ & $-0.196^{*}$  \\
\textbf{Sex} & & & & & \\
Male & Ref. & Ref. & Ref. & Ref. & Ref. & Ref.  \\
Female & $0.397$ & $0.650$ & $-3.530^{**,}\text{\textasciicircum\textasciicircum}$ & $-3.309^{*}$ & $3.073^{*}$ & $4.352^{**,}\text{\textasciicircum}$  \\
\textbf{Race} & & & & &  \\
White & Ref. & Ref. & Ref. & Ref. & Ref. & Ref. \\
Others & $3.811^{*}$ & $5.871^{**,}\text{\textasciicircum}$ & $3.137^{**,}\text{\textasciicircum}$ & $3.792^{**,}\text{\textasciicircum}$ & $1.008$ & $1.915$  \\
\textbf{BMI} & $0.674^{***,}\text{\textasciicircum\textasciicircum}$ & $0.908^{***,}\text{\textasciicircum\textasciicircum\textasciicircum}$ & $-0.021$ & $0.150$ & $-0.063$ & $0.052$ \\ 
\textbf{log(ODI4)} & $2.900$ & $2.705$ & $2.599^{**}$ & $2.554^{*}$ & $2.101$ & $1.604$ \\ 
\bottomrule
\vspace{10pt}
\end{tabular}
\caption{Coefficient estimates from multiple linear regression models with outcomes as mean daytime and nighttime SBP, DBP, and HR. The significance level before Bonferroni adjustment: *$p<0.05$, **$p<0.01$, ***$p<0.001$. The significance level after Bonferroni adjustment: \textasciicircum$p<0.05$, \textasciicircum\textasciicircum$p<0.01$, \textasciicircum\textasciicircum\textasciicircum$p<0.001$.}
\label{tab:mean_lr_res}
\end{adjustwidth}
\end{table}

\subsection{Explained variability of $L$ and $R$ components before rank reduction for parsimony}

\begin{figure}[!t]
    \centering
    \begin{subfigure}[b]{0.48\textwidth}
    \refstepcounter{subfigure}
        \begin{tikzpicture}
            \node[anchor=north west, inner sep=2pt] at (0.2,0.2) {\includegraphics[width=\textwidth]{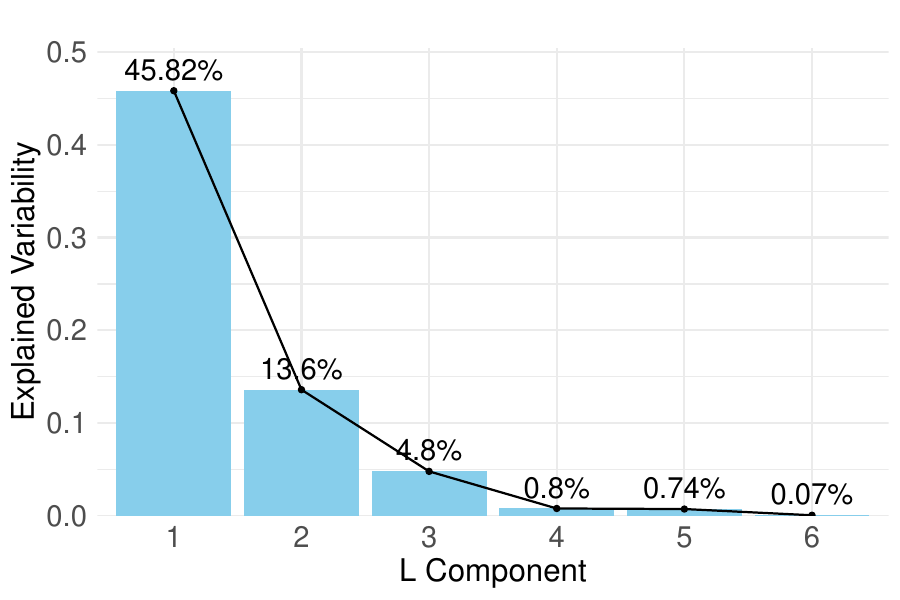}};
            \node[anchor=north west] at (0,0) {\textbf{(a)}};
        \end{tikzpicture}
        % \vspace{-35pt}
        \caption*{}
        \label{fig:var_L}
    \end{subfigure}
    \begin{subfigure}[b]{0.48\textwidth}
    \refstepcounter{subfigure}
        \begin{tikzpicture}
            \node[anchor=north west, inner sep=2pt] at (0.2,0.2) {\includegraphics[width=\textwidth]{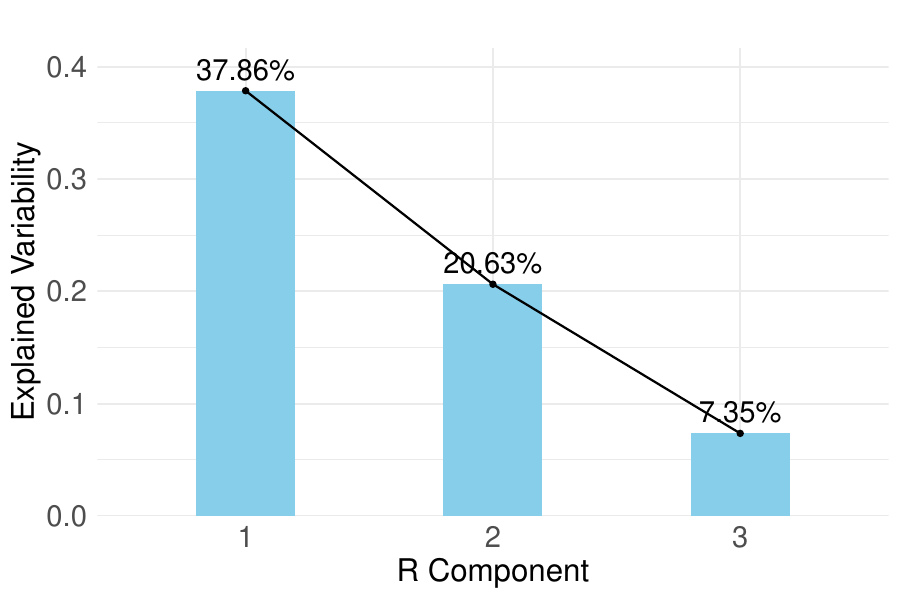}};
            \node[anchor=north west] at (0,0) {\textbf{(b)}};
        \end{tikzpicture}
        \caption*{}
        \label{fig:var_R}
    \end{subfigure}
    \vspace{-25pt}
    \caption{(a): Explained variability by the top 6 temporal components of L, when $r_2$ is fixed to 3, $\lambda$ is fixed to 12; (b): Explained variability by the 3 temporal components of R, when $r_1$ is fixed to 6, $\lambda$ is fixed to 12.}
    \label{fig:exp_var}
\end{figure}

Figure~\ref{fig:exp_var} shows the percentage of variability in the data explained by each component of the identifiability-corrected $L$ and $R$, which were generated using optimal hyperparameters from five-fold cross-validation.

\subsection{Regression models with scores for the second temporal component}

Table~\ref{tab:g2_table} illustrates linear regression model results with scores corresponding to second temporal component (nocturnal dipping). None of the baseline covariates are significant.

\begin{table}[!t]
    \centering
    \begin{tabular}{lcccc}
    \toprule
    & \multicolumn{2}{c} {\textbf{Model $g_{21}$}} & \multicolumn{2}{c} {\textbf{Model $g_{22}$}}  \\
    & \textbf{$\boldsymbol{\widehat{\beta}^{(3)}}$} & \textbf{95\% CI of $\boldsymbol{\widehat{\beta}^{(3)}}$} & \textbf{$\boldsymbol{\widehat{\beta}^{(4)}}$} & \textbf{95\% CI of $\boldsymbol{\widehat{\beta}^{(4)}}$}  \\
    \midrule
    \textbf{Age} & $-0.024$ & $(-0.058, 0.009)$ &  $0.005$ & $(-0.015, 0.026)$  \\
    \textbf{Sex} & & \\
    Male & Ref. & Ref. & Ref. & Ref. \\
    Female & $-0.452$ & $(-1.098, 0.194)$ & $-0.225$ & $(-0.620, 0.169)$  \\
    \textbf{Race} & & \\
    White & Ref. & Ref. & Ref. & Ref. \\
    Others & $0.031$ & $(-0.590, 0.651)$ & $0.210$ & $(-0.168, 0.589)$  \\
    \textbf{BMI} & $-0.017$ & $(-0.081, 0.047)$ & $0.009$ & $(-0.030, 0.048)$ \\ 
    \textbf{log(ODI4)} & $-0.267$ & $(-0.781, 0.246)$ & $0.173$ & $(-0.140, 0.487)$ \\ 
    \bottomrule
    \vspace{10pt}
    \end{tabular}
    \caption{Coefficient estimates from the multiple linear regression models, along with the corresponding 95\% confidence intervals (CIs). Model $g_{21}$: The outcome is the joint magnitude of nocturnal dipping of blood pressure and heart rate (37/100 DBP + 33/100 SBP + 3/10 HR). Model $g_{22}$: The outcome is the measure of difference between the nocturnal patterns of blood pressure and heart rate (-13/100 DBP - 33/100 SBP + 27/50 HR).}
    % The significance level before Bonferroni adjustments: *$p<0.05$, **$p<0.01$, ***$p<0.001$. The significance level after Bonferroni adjustments: \textasciicircum$p<0.05$, \textasciicircum\textasciicircum$p<0.01$, \textasciicircum\textasciicircum\textasciicircum$p<0.001$. }
    
    \label{tab:g2_table}
    % \end{adjustwidth}
\end{table}

\subsection{Interpretation of covariates from regression models on scores from first temporal component}

To further aid interpretation in terms of each physiological measurement separately on the original scale, we used the coefficient estimates for each covariate $x$ to get the effect size
$$\widehat{\Gamma}_x=(\widehat{\beta}^{(1)}_x L_1R_1^\top + \widehat{\beta}^{(2)}_x L_1R_2^\top)S,$$
where $\widehat{\beta}^{(1)}_x$ and $\widehat{\beta}^{(2)}_x$ are from the corresponding models on the scores, and $S$ is a diagonal matrix that rescales the output to original measurement scale.
Each column of $\widehat{\Gamma}_x$ corresponds to the estimated effect of $x$ on DBP, SBP, and HR. Since the scores are orthogonal, standard errors for the effect size estimates can be directly calculated as $\sqrt{\mbox{var}(\widehat{\beta}^{(1)})+ \mbox{var}(\widehat{\beta}^{(2)})}$, with subsequent rescaling by $L_1R_1^\top$, $L_1R_2^\top$, and $S$. 

Furthermore, we estimated DBP, SBP, and HR profiles by substituting $g_{11}$ and $g_{12}$ in 
\begin{equation*} \label{eq: reconstruct}
    \widehat M_i = g_{11,i}L_1R_1^\top + g_{12,i}L_1R_2^\top +
    g_{21,i}L_2R_1^\top + g_{22,i}L_2R_2^\top +
    g_{31,i}L_3R_1^\top + g_{32,i}L_3R_2^\top 
\end{equation*}
with their estimates from Model 1 and 2, and using $\widehat{g}_{21}, \ldots, \widehat{g}_{32}$ as their mean values across 191 patients obtained from algorithm implementation. 

\begin{figure}[!t]
    \centering
    \begin{subfigure}[b]{0.75\textwidth}
    \refstepcounter{subfigure}
        \begin{tikzpicture}
            \node[anchor=north west, inner sep=2pt] at (-0.1,-0.1) {\includegraphics[width=\textwidth]{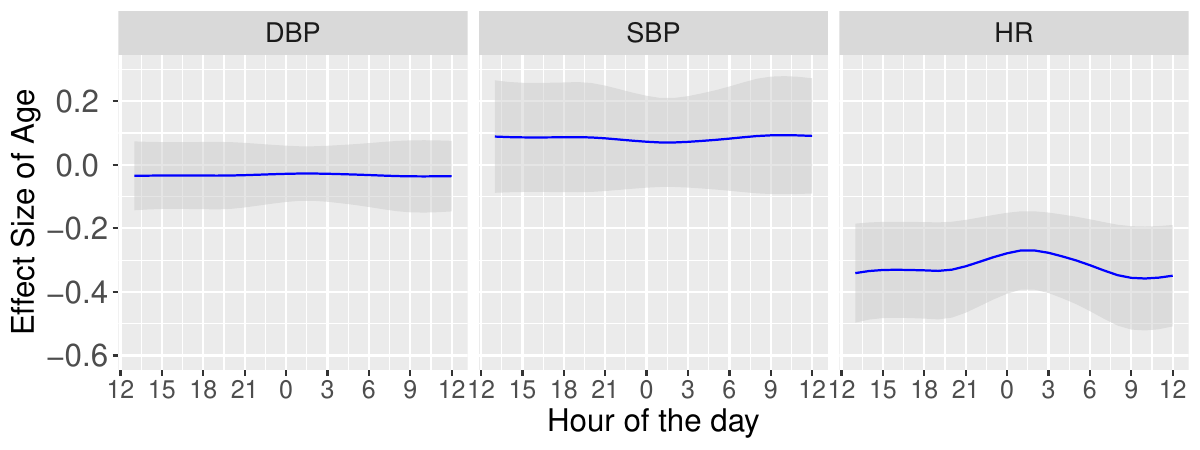}};
            \node[anchor=north west] at (0,0) {\textbf{(a)}};
        \end{tikzpicture}
         \vspace{-35pt}
        \caption*{}
        \label{fig:age_effect}
    \end{subfigure}
    \begin{subfigure}[b]{0.75\textwidth}
    \refstepcounter{subfigure}
        \begin{tikzpicture}
            \node[anchor=north west, inner sep=2pt] at (-0.1,-0.1) {\includegraphics[width=\textwidth]{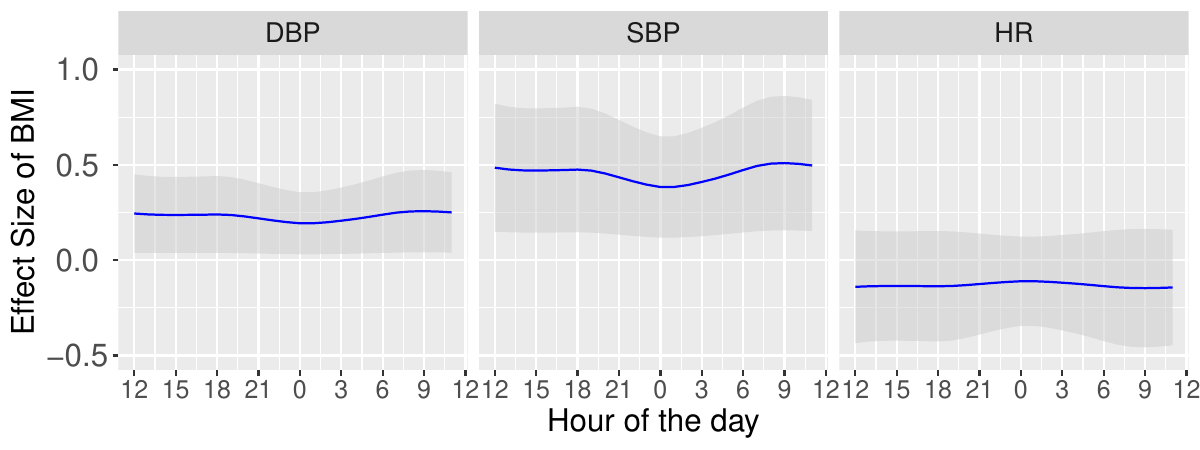}};
            \node[anchor=north west] at (0,0) {\textbf{(b)}};
        \end{tikzpicture}
        \vspace{-35pt}
        \caption*{}
        \label{fig:bmi_effect}
    \end{subfigure}
    \begin{subfigure}[b]{0.75\textwidth}
    \refstepcounter{subfigure}
        \begin{tikzpicture}
            \node[anchor=north west, inner sep=2pt] at (-0.1,-0.1) {\includegraphics[width=\textwidth]{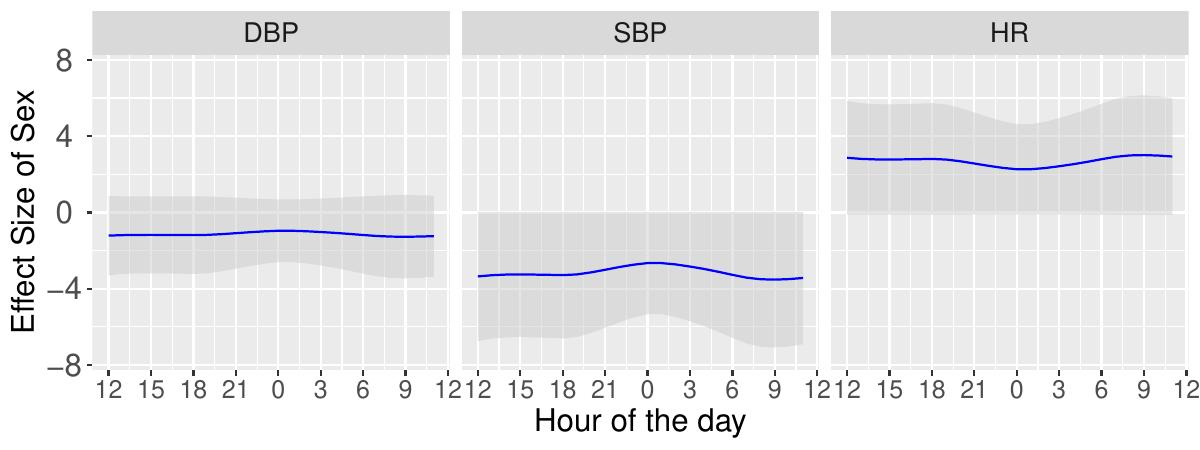}};
            \node[anchor=north west] at (0,0) {\textbf{(c)}};
        \end{tikzpicture}
        \vspace{-35pt}
        \caption*{}
        \label{fig:sex_effect}
    \end{subfigure}
    \begin{subfigure}[b]{0.75\textwidth}
    \refstepcounter{subfigure}
        \begin{tikzpicture}
            \node[anchor=north west, inner sep=2pt] at (-0.1,-0.1) {\includegraphics[width=\textwidth]{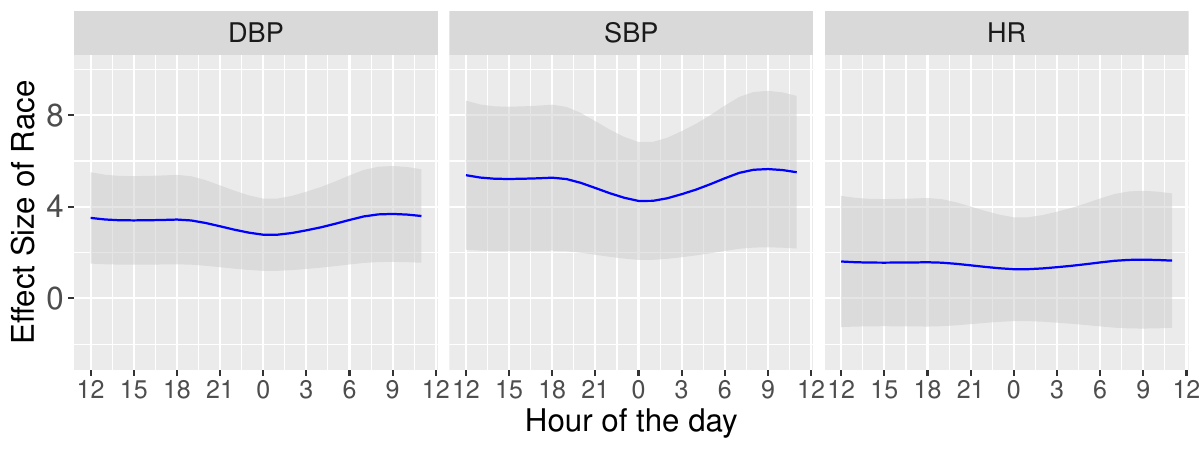}};
            \node[anchor=north west] at (0,0) {\textbf{(d)}};
        \end{tikzpicture}
        \caption*{}
        \label{fig:race_effect}
    \end{subfigure}
    
    \vspace{-25pt}
    \caption{(a): The estimated effect size of age (years) on DBP, SBP, and HR; (b) The estimated effect size of BMI ($\text{kg/m}^2$); (c): The effect size of sex being female, compared to being male; (d): The effect size of race being non-White, compared to race being White. Q1: first quartile; Q3: third quartile.}
    \label{fig:effect_plot_others}
\end{figure}

\begin{figure}[!t]
    \centering
    \begin{subfigure}[b]{0.75\textwidth}
    \refstepcounter{subfigure}
        \begin{tikzpicture}
            \node[anchor=north west, inner sep=2pt] at (-0.1,-0.1) {\includegraphics[width=\textwidth]{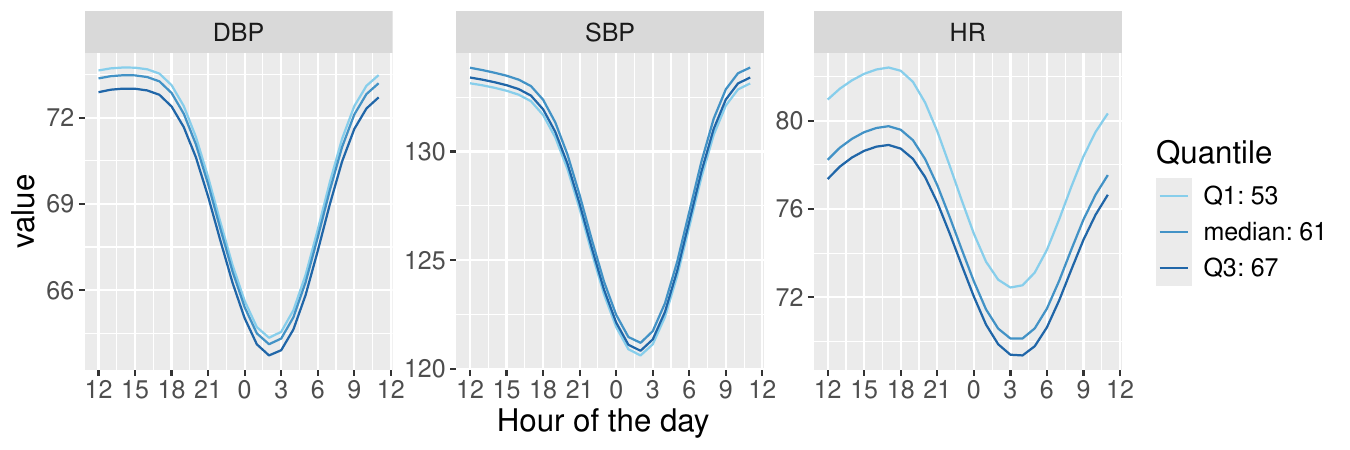}};
            \node[anchor=north west] at (0,0) {\textbf{(a)}};
        \end{tikzpicture}
         \vspace{-35pt}
        \caption*{}
        \label{fig:age_change}
    \end{subfigure}
    \begin{subfigure}[b]{0.75\textwidth}
    \refstepcounter{subfigure}
        \begin{tikzpicture}
            \node[anchor=north west, inner sep=2pt] at (-0.1,-0.1) {\includegraphics[width=\textwidth]{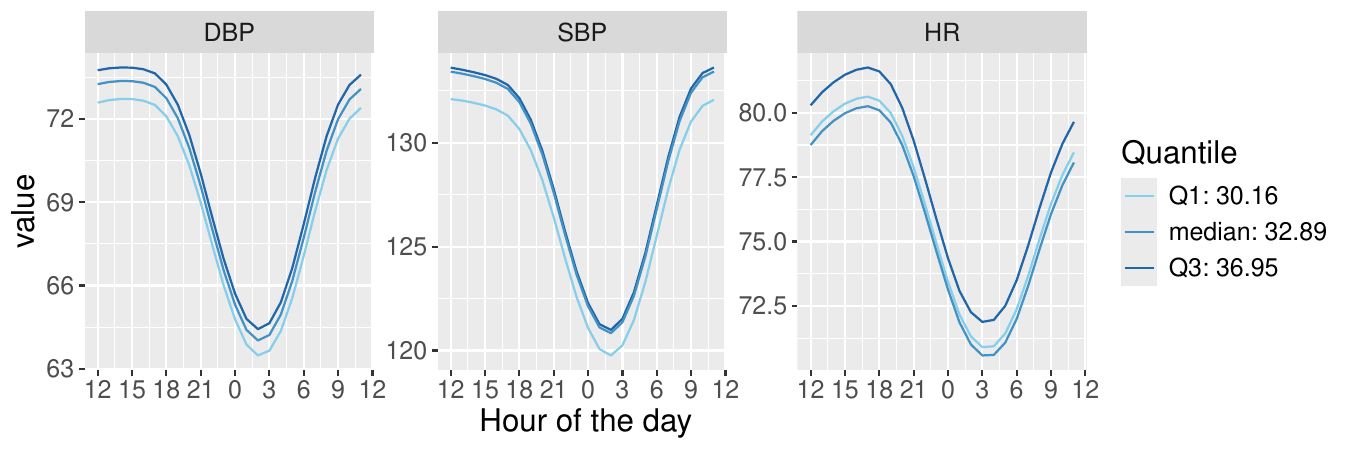}};
            \node[anchor=north west] at (0,0) {\textbf{(b)}};
        \end{tikzpicture}
        \vspace{-35pt}
        \caption*{}
        \label{fig:bmi_change}
    \end{subfigure}
    \begin{subfigure}[b]{0.75\textwidth}
    \refstepcounter{subfigure}
        \begin{tikzpicture}
            \node[anchor=north west, inner sep=2pt] at (-0.1,-0.1) {\includegraphics[width=\textwidth]{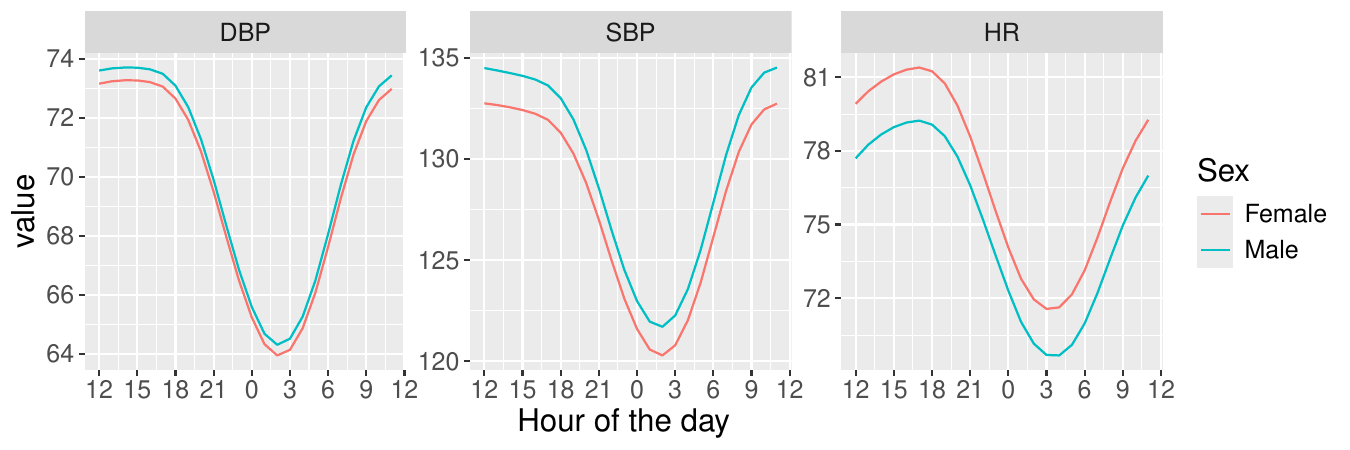}};
            \node[anchor=north west] at (0,0) {\textbf{(c)}};
        \end{tikzpicture}
        \vspace{-35pt}
        \caption*{}
        \label{fig:sex_change}
    \end{subfigure}
    \begin{subfigure}[b]{0.75\textwidth}
    \refstepcounter{subfigure}
        \begin{tikzpicture}
            \node[anchor=north west, inner sep=2pt] at (-0.1,-0.1) {\includegraphics[width=\textwidth]{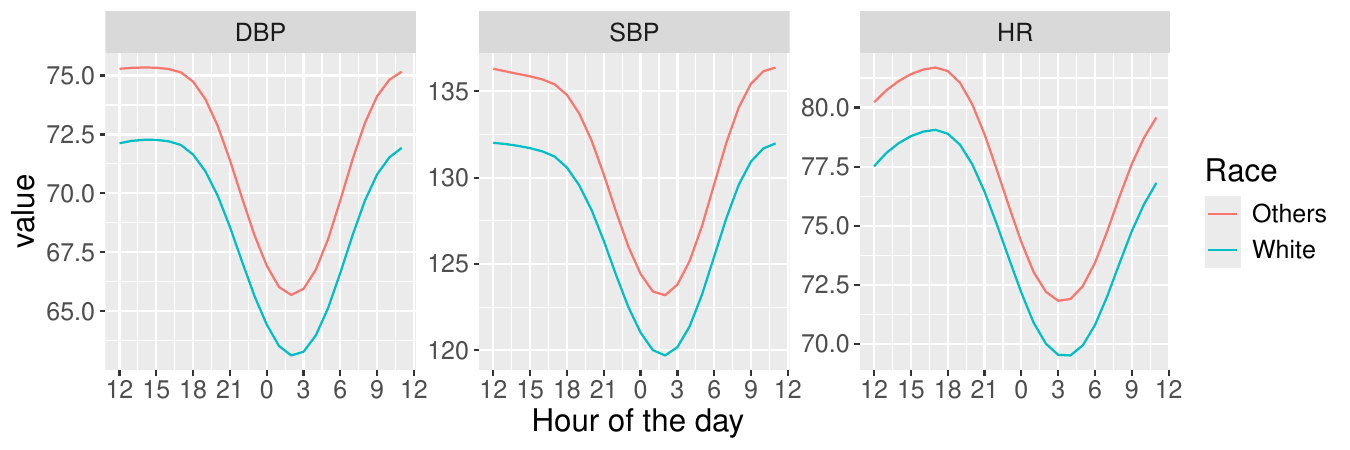}};
            \node[anchor=north west] at (0,0) {\textbf{(d)}};
        \end{tikzpicture}
        \caption*{}
        \label{fig:race_change}
    \end{subfigure}
    
    \vspace{-25pt}
    \caption{(a): The estimated effect size of age (years) on DBP, SBP, and HR; (b) The estimated effect size of BMI ($\text{kg/m}^2$); (c): The effect size of sex being female, compared to being male; (d): The effect size of race being non-White, compared to race being White. Q1: first quartile; Q3: third quartile.}
    \label{fig:change_plot_others}
\end{figure}

Figure~\ref{fig:effect_plot_others} illustrates the estimated effect sizes of each baseline covariate on the overall levels captured by the first temporal component in $L$. Figure~\ref{fig:change_plot_others} shows the predicted temporal trends of DBP, SBP, and HR, evaluated when one covariate is changed while the others are fixed at their mean levels.

\end{document}